\documentclass[twocolumn,superscriptaddress,aps,pra]{revtex4-1}
\usepackage[utf8]{inputenc}
\usepackage[T1]{fontenc}
\usepackage{newtxtext,newtxmath}
\usepackage{graphicx}
\usepackage{hyperref}
\usepackage{bm}
\usepackage{tikz}
\usepackage{microtype}
\usepackage{physics}
\usepackage{xcolor}
\usepackage{longtable}
\usepackage[caption=false]{subfig}

\begin{document}
\title{Spin-flip induced superfluidity in a ring of spinful hard-core bosons}
\author{K.~K.~Kesharpu}
\email{kesharpu@theor.jinr.ru}
\author{E.~A.~Kochetov}
\affiliation{Bogoliubov Laboratory of Theoretical Physics, Joint Institute for Nuclear Research, Dubna 141980, Russia}
\date{\today}
\begin{abstract}
   The $t-J$ Hamiltonian of the spinful hard-core bosonic ring in the \emph{Nagaoka limit} is solved. The energy spectrum becomes quantized due to presence of spin, where each energy level corresponds to a cyclic permutation state of the spin chains. The ground state is true ferromagnetic when the ring contains $N=2,3$ spinful hard-core bosons; for all other $N$ it is a mixture of the ferromagnetic and non-ferromagnetic states. This behaviour is different from the fermionic ring, where ground state is true ferromagnetic only for $N=3$. It is shown that the intrinsic spin generated gauge fields are analogous to the synthetic gauge fields generated by rotation of either the condensate or the confining potential. It is argued that the low lying excited levels of the spin flipped states intrinsically support the superfluidity. Possible ways to experimentally verify these results are also discussed.
\end{abstract}

\maketitle

\section{Introduction}
\label{sec:introduction}

Rapid progress in the experimental techniques of trapping and manipulating cold atoms has opened numerous possibilities for their use in quantum simulation and quantum computing \cite{georgescu-QuantumSimulation-2014,bruzewicz-TrappedionQuantumComputing-2019,schafer-ToolsQuantumSimulation-2020,blochUltracoldQuantumGases2005}. Spinor bosons---atoms with integer spins---in optical lattice are one of the modern work horse to probe the physics of the strongly correlated systems \cite{bloch-QuantumSimulationsUltracold-2012,bloch-QuantumSimulationsUltracold-2012,blochManybodyPhysicsUltracold2008,lewensteinUltracoldAtomicGases2007}. It has several advantages over their condensed matter counterparts e.g., precise knowledge of underlying microscopic model, possibility to control the parameters of lattice Hamiltonian, and absence of impurity in physical realizations. In this work we theoretically investigate the physics of one of the simplest system, yet rich in physics, that can be constructed using spinor bosons: the \emph{1D} ring lattice loaded with spinful hard-core bosons (HCB) \footnote{When two or more bosons can not occupy the same position due to strong repulsive inter particle potential, they are known as hard-core bosons \cite{girardeau-RelationshipSystemsImpenetrable-1960}.}.

The spinless bosonic ring is a well studied problem both theoretically and experimentally \cite{morsch-DynamicsBoseEinsteinCondensates-2006,cazalilla-OneDimensionalBosons-2011,manninen-QuantumRingsBeginners-2012,blochManybodyPhysicsUltracold2008,lewensteinUltracoldAtomicGases2007}. In Ref. \cite{manninen-QuantumRingsBeginners-2012} the \emph{Yarst} states for fermionic and bosonic ring were given. In Ref. \cite{cazalilla-OneDimensionalBosons-2011} several exactly solvable models for one dimensional bosonic systems were reviewed. In Ref. \cite{wang-SpontaneousAharonovCasherEffect-1995} a model for neutral \emph{spinless} HCB on a ring was solved. Ref. \cite{boninsegni-PhaseSeparationMixtures-2001} considered \emph{spinful} HCB on a $N \times N$ plaquette. On this plaquette they solved the usual \emph{t-J} Hamiltonian by dividing the spinful HCB into two different species of spinless HCB. Ref. \cite{fledderjohann-FerromagnetismHardcoreBoson-2005} also considered spinful HCB on finite two-dimensional plaquette. They investigated the effect of fermionic and bosonic statistics on the emergence of ferromagnetic phase. In Ref. \cite{bergkvistSpinfulBosonsOptical2006} the phase diagram of \emph{1D} chain, when even and odd number of spinful bosons are present at a single site, were investigated. In Ref. \cite{nie-ParticleStatisticsFrustration-2018,nie-GroundStateEnergiesSpinless-2013} the ground state properties of the spinful fermions and bosons in thermodynamic limits were studied. Most of the work done till now are related either to the bosons on one and two dimensional periodic lattices, and their behaviour in the thermodynamic limit ($N \gg 1$), or \emph{spinless} HCB on a ring. Apart from some comments about the energy levels of spinful bosonic ring in Ref. \cite{manninen-QuantumRingsBeginners-2012}, a comprehensive study of the properties of the spinful hard-core bosons on a ring away from the thermodynamic limit is still missing.

One of the interesting effect observed in the ring of bosons is the persistent current, which is related to the superfluidity \cite{cooperRapidlyRotatingAtomic2008,fetterRotatingTrappedBoseEinstein2009,vilchynskyyNatureSuperfluidityBoseEinstein2013}. In Ref. \cite{caiSuperfluidtoBoseglassTransitionHardcore2010} the phase diagram of superfluid and insulating phases was studied for spinless hard-core bosons. Ref. \cite{linGroundStateHardcore2007} investigated the ground state and superfluidic properties of the spinless hard-core bosons in one dimensional potential. Recently, Ref. \cite{poloExactResultsPersistent2020} estimated the values of persistent current for two hard-core bosons in a ring lattice. From the experimental side the persistent current was observed in spinor (not hard-core) condensates \cite{beattiePersistentCurrentsSpinor2013}, and fermion rings \cite{chinEvidenceSuperfluidityUltracold2006,caiPersistentCurrentsRings2022}. It is well known that ground state can never support the persistent current \cite{pitaevskiiBoseEinsteinCondensationSuperfluidity2016,liebSuperfluidityDiluteTrapped2002,leggettSuperfluidity1999}. Hence, one accesses the excited superfluidic states by applying a velocity field, either by rotating the confined particles in the ring lattice, or the ring lattice itself \cite{liebSuperfluidityDiluteTrapped2002,cooperRapidlyRotatingAtomic2008,fetterRotatingTrappedBoseEinstein2009,vilchynskyyNatureSuperfluidityBoseEinstein2013,pitaevskiiBoseEinsteinCondensationSuperfluidity2016}. The rotation of the lattice is analogous to the generation of synthetic gauge fields, which in turn is related to the twisted boundary condition \cite{liebSuperfluidityDiluteTrapped2002,cooperRapidlyRotatingAtomic2008}. We show that, for the case of hard-core spinful bosons the twisted boundary condition is generated intrinsically without application of any external velocity field.

\emph{The main focus of our work is the investigation of the ground state properties and the necessary condition for occurrence of superfluidity in these systems.} This article is structured as follows. In Sec. \ref{sec:hamilt-spinf-hcb} we solve the Hamiltonian of the spinful HCB on a ring in \emph{Nagaoka limit}. In this limit the spin and charge degrees of freedom can be treated separately, hence, the Hamiltonian is easily solvable. In Sec. \ref{sec:ground-state-prop-1} we investigate the dependence of the ground state energies on the total spin and the structure of the total spin chains. We discuss the necessary conditions for the emergence of superfluidity in these systems. Here, we also suggest the experimental setups to corroborate our theoretical predictions. Finally, In Sec. \ref{sec:conclusion-1} we summarise the results.

\section{The Model}
\label{sec:hamilt-spinf-hcb}

We take a ring of $L$ sites and $N$ spinful HCB with spin projections $\sigma=\left\{ \uparrow, \downarrow \right\}$ \footnote{Although, the bosons don't have two spin projections $\pm 1/2$ like fermions, but two \emph{hyperfine} states of some atoms can be considered as two spin projections of the bosons. These spins are known as the \emph{pseudo spins}, and these atoms are known as the \emph{spinor bosons} \cite{stamper-kurn-SpinorBoseGases-2013,kawaguchi-SpinorBoseEinstein-2012}. Naturally, one can map the \emph{pseudo spins} to the spin-up state $(\ket{\uparrow})$ and the spin-down state $(\ket{\downarrow})$.}. Because of the hard-core nature of the bosons, in the ring every site contains \emph{at most single boson} ($N \leq L$). The \emph{t-J} Hamiltonian in the \emph{Nagaoka limit} with periodic boundary conditions \cite{nagaoka-FerromagnetismNarrowAlmost-1966,nagaoka-GroundStateCorrelated-1965} can be written as [see supplementary materials]:
\begin{equation}
  \label{eq:Ham-with-spin}
  H = - t \sum\limits_{i=1, \sigma}^{L-1} \tilde{b}_{i \sigma}^{\dagger} \tilde{b}_{i+1 \sigma} - t \sum\limits_{\sigma} \tilde{b}_{L \sigma}^{\dagger} \tilde{b}_{1 \sigma} + \mathrm{H.c}.
\end{equation}
Here, $t$ is the boson hopping factor from the \emph{i}-th site to the neighbouring \emph{i+1}-th site. $\tilde{b}_{i \sigma}^{\dagger}$ and $\tilde{b}_{i \sigma}$ are the HCB creation and annihilation operators with spin projections $\sigma=\left\{ \uparrow, \downarrow \right\}$. The operators $\tilde{b}_{i \sigma}$ and $\tilde{b}_{i \sigma}^{\dagger}$ implies single occupancy constraint: $\sum\limits_{\sigma} b_{i \sigma}^{\dagger} b_{i \sigma}\leq 1$.

Due to the 1D nature of the ring, and the absence of the spin interaction, the initial spin configuration of the spin chain is fixed during electron hopping. Hence, we can separate the spin and charge degree of freedoms in the Hamiltonian:
\begin{equation}
  \label{eq:Ham-hcb}
  H = -t \sum\limits_{i=1}^{L-1} {b}_{i}^{\dagger} {b}_{i+1} -t {b}_{L}^{\dagger} {b}_{1} \hat{P} + \mathrm{H.c}.
\end{equation}
Here, ${b}_{i}^{\dagger} ({b}_{i})$ is a \emph{spinless} hard-core bosonic creation (annihilation) operator. 
The spin content of the problem is encoded in 
the spin permutation operator $\hat{P}$. It displaces the spin to the next non-empty site:
\begin{equation*}
  \label{eq:spin-permut}
  \hat{P} \ket{s_{1}^{z}, s_{2}^{z}, \dots, s^{z}_{N}} = \ket{s_{N}^{z}, s_{1}^{z}, \dots, s^{z}_{N-1}}.
\end{equation*}
The boson hopping part of Eq. (\ref{eq:Ham-hcb}) is analogous to the \emph{XY} spin-chain Hamiltonian \cite{holstein-FieldDependenceIntrinsic-1940,cazalilla-OneDimensionalBosons-2011,lieb-TwoSolubleModels-1961,wang-SpontaneousAharonovCasherEffect-1995}. Hence, one can use the \emph{Jordan-Wigner transformation} to represent the Hamiltonian in Eq. (\ref{eq:Ham-hcb}) in terms of spinless fermionic operators [see supplementary materials]:
\begin{equation}
  \label{eq:Hamiltonian-ferm-opr-period}
  H = - t \sum\limits_{i=1}^{L-1}\hat{f}_i^{\dagger}\hat{f}_{i+1} + \: t \: \mathrm{e}^{\imath \pi N} \hat{f}_{L}^{\dagger}\hat{f}_1\: \hat{P}+ \text{H.c}.
\end{equation}
Here, $\hat{f}_i^{\dagger}$ ($\hat{f}_i$) is the spinless fermionic creation (annihilation) operator. Defining the function,
\begin{equation}
  \label{eq:Hn-funct}
  h(N) =
  \begin{cases}
    0, \quad \text{odd }N\\
    1, \quad \text{even }N
  \end{cases}
  ,
\end{equation}
we can write Eq. (\ref{eq:Hamiltonian-ferm-opr-period}) as:
\begin{equation}
  \label{eq:Ham-ferm-period-odd-even}
    H = - t \sum\limits_{i=1}^{L-1}\hat{f}_i^{\dagger}\hat{f}_{i+1} - \: t \: \mathrm{e}^{\imath \pi h(N)} \hat{f}_{L}^{\dagger}\hat{f}_1\: \hat{P}+ \text{H.c}.
\end{equation}

The spin permutation and spinless fermionic operators are separately diagonalized, because they are independent of each other. The eigenvalues ($\lambda_{\nu}$) and eigenfunction ($\psi_{\nu}$) of $\hat{P}$ are \cite{ivantsov-StableMetastableKinetic-2020}:
\begin{subequations}
\begin{equation}
  \label{eq:spin-eigen-value}
  \begin{aligned}
    &\lambda_{\nu} = \mathrm{e}^{\imath 2 \pi p_{\nu}/ N_{\nu}};\\
  \end{aligned}
\end{equation}
\begin{equation}
  \label{eq:spin-eigen-fun}
  \begin{aligned}
    &\ket{\psi_{\nu}} = \frac{1}{\sqrt{N_{\nu}}} \sum\limits_{q=0}^{N_{\nu}-1} \mathrm{e}^{\imath 2 \pi p_{\nu} \frac{q}{N_{\nu}}} \hat{P}^q \: \ket{ \tilde{\psi}_{\nu}}.
  \end{aligned}
\end{equation}
\end{subequations}
Here, $\nu$ enumerates all possible disconnected spin blocks. A spin block contains only connected spin chains. When two spin chains can be transformed into each other by application of $\hat{P}$ operator they are connected, otherwise they are disconnected. $N_{\nu}$ represents the total number of connected spin chains in $\nu$-th spin block. $p_{\nu}$ enumerates the connected spin chains in the $\nu$-th spin block; it takes the values $p_{\nu}= 0, 1, \dots, N_{\nu}-1$. $\tilde{\psi}_{\nu}$ is the wave function of one of the spin chain of $\nu$-th spin block.

For example, we have a chain of 4 sites and 3 particles. We take the spin-chain $\ket{\uparrow \bullet \uparrow \downarrow}$ out of $2^{3}$ possible spin-chains. It is connected to the $\ket{\uparrow \bullet \downarrow \uparrow}$ spin-chain, as $\hat{P}^{2} \ket{\uparrow \bullet \uparrow \downarrow} = \ket{\uparrow \bullet \downarrow \uparrow}$. In this case both these configurations belong to the same $\nu$-th spin block. This particular $\nu$-th block will have 3 possible configurations, hence $N_{\nu}$ = 3, and $p_{\nu} = 0, 1, 2$ \footnote{The three configurations ($N_{\nu}$ = 3) corresponding to three $p_{\nu}$ are: (i) $p_{\nu}=0 \equiv \ket{\uparrow \bullet \uparrow \downarrow}$, (ii) $p_{\nu}=1 \equiv \ket{\downarrow \bullet \uparrow \uparrow }$, (iii) $p_{\nu}=2 \equiv \ket{\uparrow \bullet \downarrow \uparrow}$.}. The wave function of three $p_{\nu}$ states can be found using Eq. (\ref{eq:spin-eigen-fun}) [see App. \ref{sec:matr-repr-spin}]. Consequently, every spin-chain in the $\nu$-th spin block has its own wave function and spin momentum $p_{\nu}$. The number of disconnected blocks will depend on the number of particles present in the ring ($N$) and the spin of these particles ($s^{z}$). Due to these disconnected block of spin chains, the total spin Hamiltonian corresponding to the $\hat{P}$ operator is a block Hamiltonian with $\nu$ blocks. We find the Hamiltonian corresponding to the $\nu$-th block by substituting $\lambda_{\nu}$ from Eq. (\ref{eq:spin-eigen-fun}) into Eq. (\ref{eq:Ham-ferm-period-odd-even}):
\begin{equation}
  \label{eq:ham-spin-opr}
  H_{\nu} =
    - t \sum\limits_{i=1}^{L-1}\hat{f}_i^{\dagger}\hat{f}_{i+1} - \: t \: \mathrm{e}^{i 2 \pi \left[ \frac{p_{\nu}}{N_{\nu}} + \frac{h(N)}{2}\right]} \hat{f}_{L}^{\dagger}\hat{f}_1 + \text{H.c}.
\end{equation}
The total Hamiltonian of the whole system is a direct sum of these spin block Hamiltonians: $H = \sum\limits_{\nu} \oplus H_{\nu}$.

Eq. \eqref{eq:ham-spin-opr} is nothing but the tight binding model with a penetrating magnetic flux $\Phi_{\nu}\equiv 2 \pi \left[ \frac{p_{\nu}}{N_{\nu}} + \frac{h(N)}{2}\right]$ through the ring. Using the gauge $f_{i} \mapsto \mathrm{e}^{\imath \Phi_{\nu} x_{i}/L} f_{i}$ one maps the Hamiltonian in Eq. (\ref{eq:ham-spin-opr}) onto twisted Hamiltonian:
\begin{equation}
  \label{eq:ham-gauge}
  H_{\nu}=
      - t \sum\limits_{i}^{L-1}\mathrm{e}^{\imath \frac{\Phi_{\nu}}{L}}\hat{f}_i^{\dagger}\hat{f}_{i+1}
      - \: t \: \mathrm{e}^{\imath \frac{\Phi_{\nu}}{L}} \hat{f}_{L}^{\dagger}\hat{f}_1 + \text{H.c}.
\end{equation}
Here, $x_{i} = 1, 2, \dots, L$, enumerates the $L$ sites. The locally induced phase factor $\mathrm{e}^{\imath \Phi_{\nu}/L}$ is known as \emph{Peierls phase}.
The explicit expression for the $k$-th mode energy of this tight binding Hamiltoinian $H_{\nu}$ is [see supplementary materials]:
\begin{equation}
  \label{eq:eigen-energy-pbc}
  \begin{aligned}
  E_{\text{PBC}}(k, \nu, p_{\nu}; N, L)=
  -2t \cos \frac{2 \pi}{L}\left( k+ \frac{p_{\nu}}{N_{\nu}} + \frac{h(N)}{2} \right).
  \end{aligned}
\end{equation}
The total energy is found by summing over all $N$ low lying $k$-th mode energies:
\begin{widetext}
  \begin{equation}
    \label{eq:app:Eg:odd-N-PBC-wo-mag}
    \begin{aligned}
      E_{PBC, \: g} =& -2t\frac{\sin \left[ (N+1+h(N))\pi/2L \right]}{\sin (\pi/L)} \cos \left[ \frac{\frac{4\pi}{L}\left( \frac{p_{\nu}}{N_{\nu}} + \frac{h(N)}{2} \right) + \frac{(N-1+h(N))\pi}{L}}{2} \right]\\ & -2t\frac{ \sin\left[ (N+1+h(N))\pi/2L \right]}{\sin (\pi/L)} \cos \left[ \frac{\frac{4\pi}{L}\left( \frac{p_{\nu}}{N_{\nu}}  + \frac{h(N)}{2} \right) - \frac{(N-1+h(N))\pi}{L}}{2} \right]\\ &+ 2t \left[ 1+h(N) \right] \cos \left[ \frac{2 \pi}{L} \left( \frac{p_{\nu}}{N_{\nu}} + \frac{h(N)}{2}  \right) \right].
      \end{aligned}
    \end{equation}
\end{widetext}

The Hamiltonian of the antiperiodic boundary condition $\left(\tilde{b}_{L+1, \sigma} = -\tilde{b}_{1, \sigma}\right)$ is written by introducing the extra phase $\mathrm{e}^{i \pi}$ in the second term of Eq. (\ref{eq:Ham-with-spin}). Repeating the aforementioned procedure, the \emph{k}-th mode energy levels and the ground state energies can be calculated. One can directly find these expressions from Eq. (\ref{eq:eigen-energy-pbc}) and (\ref{eq:app:Eg:odd-N-PBC-wo-mag}) by replacing $h(N) \to h(N+1)$. If a magnetic field $\mathbf{B}$ is applied perpendicular to the ring, then an additional flux $\Phi_{B} = 2 \pi \mathbf{B} A$---here, $A$ is the area of the ring---penetrates through the ring. To find the total energy one repeats the above calculation by substituting $\Phi_{\nu} \mapsto \Phi_{\nu} + \Phi_{B}$, and adds the total spin ($S_{\nu}$) dependent \emph{Zeeman energy}, $Z=g \mu_{B} B S_{\nu}$ [see supplementary materials]. Here, $g$ is the \emph{Lande factor}; $\mu_{B}$ is the \emph{Bohr magneton}; $S_{\nu}$ is the total spin of the $\nu$-th block.

\section{Ground state properties and Superfluidity}
\label{sec:ground-state-prop-1}

\begin{figure}[tbh]
  \centering
  \includegraphics[width=0.5\textwidth]{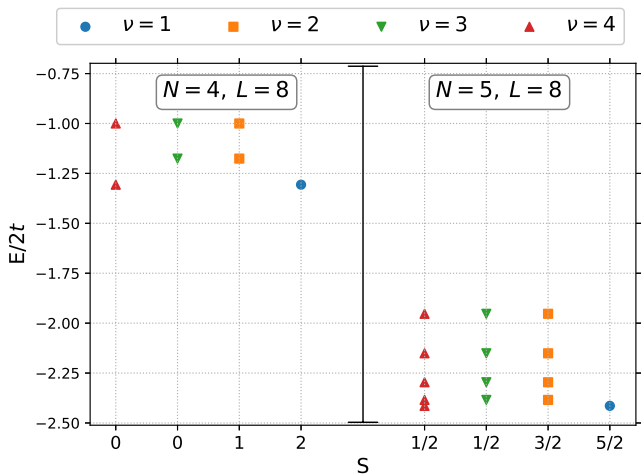}
  \caption{(Color online) Energy levels for the periodic boundary condition when  $N=4,5$ bosons reside on a ring of $L=8$ sites. The \emph{x}-axis represents the spin ($\mathrm{S}$), and the \emph{y}-axis represents the energy ($E/2t$). $\nu$ represents the spin block. The spin chain configurations corresponding to different $\mathrm{S}$ and $\nu$ are shown in Tab. \ref{tab:app:spin-config-n4l6} and \ref{tab:app:spin-config-n5l6}. For $\nu=2$ (square, orange) and $\nu=3$ (down triangle, green) the $p_{\nu}=0$ state is not available [see App. \ref{app:pstate}]. For both $S=0$ and $S=1/2$ spin states, correspond two spin blocks: $\nu=3$ (down triangle, green) and $\nu=4$ (up triangle, red).}
  \label{fig:N4N5L6}
\end{figure}
In a ring states corresponding to the cyclic permutation ($p_{\nu}$) of the initial spin configurations become available, because particles can jump directly from the \emph{L}-th site to the \emph{1}-st site,  One can further group these permutation states into irreducible representations of the cyclic symmetry groups $C_{n}$. It should be noted that a single cyclic group can contain more than one spin blocks ($\nu$). For example, for $N=4$ the $C_{4}$ group contains the $\nu=2,3$ spin blocks as shown in Tab. \ref{tab:app:spin-config-n4l6}. The more general relation between $N$, $p_{\nu}$, and $\nu$ can be found using \emph{Burnside's Lemma} \cite{xavier-OnsetFerromagnetismStrongly-2020}.

We show the detailed spin configurations for $N=4$ and $5$ in Tab. \ref{tab:app:spin-config-n4l6} and \ref{tab:app:spin-config-n5l6} respectively. The corresponding energy levels with $L=8$ are shown in Fig. \ref{fig:N4N5L6}. Here, for $N \geq 4$ the ground state is a mixture of the ferromagnetic phase ($S=N/2$) and non-ferromagnetic phases ($S=N/2-2, N/2-3, \dots$). It should be noted that the single spin flipped phase ($S=N/2-1$) is absent in the ground state due to the unavailability of the $p_{\nu}=0$ state [see App. \ref{app:pstate}]. Interestingly for $N=2,3$, the ground state is pure ferromagnetic, because the spin flipped phases ($S=N/2-2, N/2-3, \dots$) are not available. This behaviour is different from the fermionic ring, for which the ground state is pure ferromagnetic only for $N=3$ \cite{ivantsov-StableMetastableKinetic-2020}.

Physically the spinful hard-core bosonic ring can be realised by loading spinor bosons \cite{stamper-kurn-SpinorBoseGases-2013,kawaguchi-SpinorBoseEinstein-2012} in optical tweezers \cite{grier-RevolutionOpticalManipulation-2003,neuman-OpticalTrapping-2004,degoerdeherveVersatileRingTrap2021} or paul traps \cite{schneider-ExperimentalQuantumSimulations-2012,tomza-ColdHybridIonatom-2019}. The two hyperfine states ($F$) of the spinor bosons can be considered as two \emph{pseudo-spin} states. Recently numerous experiments have successfully generated several 2D and 3D crystals with high fidelity \cite{nogrette-SingleAtomTrappingHolographic-2014,barredo-SyntheticThreedimensionalAtomic-2018,barredo-AtombyatomAssemblerDefectfree-2016,endres-AtombyatomAssemblyDefectfree-2016,kim-SituSingleatomArray-2016,donofrio-RadialTwoDimensionalIon-2021}. Hence generating 1D rings should not be difficult. One of the interesting fact to observe experimentally is the dependence of the energy levels on underlying spin structures. For example one can generate a ring of spinor bosons with the initial spin configuration $\ket{\uparrow \uparrow \downarrow \downarrow}$ ($\nu=3$ in Tab. \ref{tab:app:spin-config-n4l6}). Then the system is excited to the higher energy levels ($p_{\nu}=1,2,3$) through rotation of the confining potentials \cite{cooperRapidlyRotatingAtomic2008,fetterRotatingTrappedBoseEinstein2009}. To return to the ground state the system should radiate the energy proportional to the $E(p_{\nu})-E(p_{\nu}=0)$, which can be easily measured. In the next step one can prepare the system with the spin arrangement $\ket{\uparrow  \downarrow \uparrow \downarrow}$ ($\nu=4$ in Tab. \ref{tab:app:spin-config-n4l6}). Analogously the system will be excited to the higher energy levels, and the radiated energy will be measured. In the former case the radiated energy will be higher than the later case, because only single $p_{\nu}=1$ state is available. \emph{It will be the direct experimental evidence of the spin chain configuration dependent quantization of the energy in the spinful hard-core bosonic rings.}

As an example of the physics displayed by the spinful hard-core bosonic ring, let us show that a slight change of the underlying spin structure of the HCB on a ring might provide a necessary  condition for a superfluidity to emerge.
According to the two-fluid picture, the superfluid contains both normal as well as superfluid components. One therefore defines a quantity, the so called \emph{superfluid fraction ($f_{s}$)}, to represent the \emph{degree of superfluidity}. There are different ways to calculate it \cite{liebSuperfluidityDiluteTrapped2002,prokofevTwoDefinitionsSuperfluid2000}. We use the definition where $f_{s}$ is calculated through the reaction of the system under a change in boundary conditions. Mathematically, the change in boundary conditions is equivalent to imposing a linear phase variation $\Theta x/L$ over length $L$ of the system \cite{liebSuperfluidityDiluteTrapped2002}. Hence, if $\Psi(x)$ is the wave function of the superfluid, then $\Psi(x+L) = \mathrm{e}^{i \Theta} \Psi(x)$. It should be stressed that, the phase variation should be linear in $x$ to conserve the symmetry of the system and avoid a phase slip. Physically it means that the particles acquire a similar phase $\Theta/L$ while tunneling to the neighbouring sites.

\begin{figure}[tbh]
  \centering
  \includegraphics[width=0.5\textwidth]{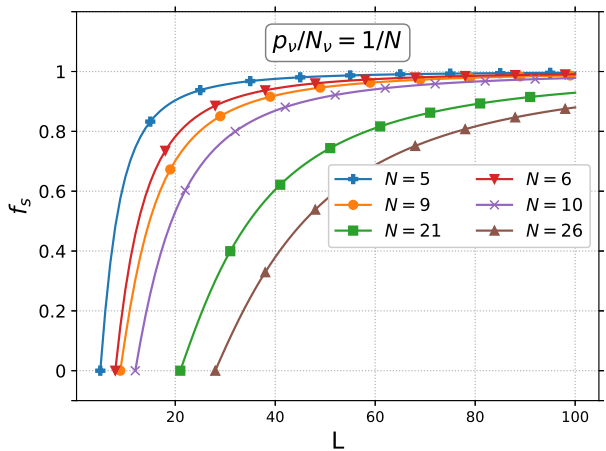}
  \caption{(Color online) Dependence of superfluid fraction ($f_{s}$) on number of sites ($L$) while number of particles ($N$) is fixed. $f_{s}$ is calculated using Eq. (\ref{eq:sc-frac}) with phase $\theta=2 \pi/N$ for $N=5$ (blue, plus), $N=6$ (red, down-triangle), $N=9$ (orange, circle), $N=10$ (violet, cross), $N=21$ (green, square), and $N=26$ (brown, up-triangle).}
  \label{fig:sup-fl-fr}
\end{figure}
Physically, the twisted phase $\Theta$ is imposed by rotating the system with some angular velocity $\omega$ \cite{liebSuperfluidityDiluteTrapped2002,rothPhaseDiagramBosonic2003,rothSuperfluidityInterferencePattern2003}. For a unit radius 1D ring, $\Theta$ is related to the superfluid velocity: $v_{s}= \hbar \Theta/(mL)$ \cite{rothSuperfluidityInterferencePattern2003,leggettSuperfluidity1999,liebSuperfluidityDiluteTrapped2002}. Experimentally, the twisted phase can be imposed through atom-light interactions \cite{dalibardColloquiumArtificialGauge2011}, rotating the confining potential \cite{cooperRapidlyRotatingAtomic2008}, or rotating the confined atoms \cite{fetterRotatingTrappedBoseEinstein2009}. However, there is another way to impose twisted boundary condition: \emph{through the change in the underlying spin configurations.} Indeed the phase factor $\Phi_{\nu}/L$ in Eq. (\ref{eq:ham-gauge}), which is dependent on the spin configuration through $\nu$, is equivalent to the twisted phase. The persistent current appears when $\Phi_{\nu} \ll \pi$, because only for this case that high energy excitation are absent in the system \cite{rothPhaseDiagramBosonic2003}. In this limit the superfluid density is \cite{rothSuperfluidityInterferencePattern2003}:
\begin{equation}
  \label{eq:sc-frac}
  f_{s} = \frac{L^{2}}{t N} \frac{E_{\Phi_{\nu}} - E_{0}}{\Phi_{\nu}^{2}}.
\end{equation}
Here, $E_{0}$ ($E_{\Phi_{\nu}}$) is the energy of the system in the absence (presence) of the phase twist.

Eq. (\ref{eq:sc-frac}) is directly applicable when the ring contains \emph{odd and large} number of particles. In this case the low lying excited energy levels ($p_{\nu} \ll N_{\nu}$) satisfy the condition $\Phi_{\nu}^{\text{odd}} := 2 \pi p_{\nu}/N_{\nu} \ll \pi$. For even $N$ the twisted phase takes on the form $\Phi_{\nu}^{\text{even}}:= 2 \pi p_{\nu}/N_{\nu} + \pi$. An extra phase factor of $\pi$ accounts for a passage from odd $N$ to even $N$. Therefore to calculate $f_{s}$ induced solely by a change in the spin structure at fixed even $N$, we should replace $\Phi_{\nu}^{\text{even}} \mapsto \Phi_{\nu}^{\text{even}}-\pi = \Phi^{\text{odd}} \ll \pi$. In Fig. \ref{fig:sup-fl-fr} we plotted the dependence of the superfluid fraction on the number of sites in the ring for $N \geq 5$ and $p_{\nu}/N_{\nu}=1/N$. It can be observed that, when $N \approx L$ (commensurate) superfluidity is absent ($f_{s} \approx 0$). However, the superfluid fraction increases as $L$ increases (incommensurate). The occurrence of the superfluidity for incommensurate case ($N/L \ll 1$) is a manifestation of the fact that a $1D$ dilute gas of hard-core bosons is always superfluid \cite{liebSuperfluidityDiluteTrapped2002}.

It should be mentioned that, the condition $\Phi_{\nu} \ll \pi$, is \emph{necessary but not sufficient} for appearance of superfluidity, because it does not say anything about the stability of the persistent currents \cite{pitaevskiiBoseEinsteinCondensationSuperfluidity2016}. We propose the following experiment  to detect superfluidity generated by a single-spin flip. One can prepare a 1D ring of spinor bosons using optical tweezers, and containing $N \gg 1$. All the particles should be in the same spin state, in other words the system is in ferromagnetic phase ($S=N/2$). Then spin of a single atom is flipped. Due to absence of the $p_{\nu}=0$ state for spin $S=N/2-1$, the ground state will then be the first excited state ($p_{\nu}=1$). It is equivalent to generating a small twisted phase $\Phi= 2\pi/N$. One can then experimentally find the \emph{matter-wave interference pattern} and \emph{structure factor} to get information about superfluidity \cite{rothPhaseDiagramBosonic2003}.

\section{Conclusions}
\label{sec:conclusion-1}

We have shown that in the spinful hard-core bosonic ring energy spectrum is quantized. Here each excited energy level corresponds to a cyclic permutation state ($p_{\nu}$) of the spin chain. Interestingly, depending on the underlying spin configuration of the spin chains, the spin blocks with identical total spin ($S$) can have different size ($N_{\nu}$). For example, although the total spin of the spin-chains $\ket{\uparrow \uparrow \downarrow \downarrow}$ and $\ket{\uparrow \downarrow \uparrow \downarrow}$ are same ($S=0$), however due to different spin configurations the size of the corresponding spin blocks will be four and two respectively [see Tab. \ref{tab:app:spin-config-n4l6}]. To corroborate this fact, one can perform experiments measuring the radiated energy from the first excited state ($p_{\nu}=1$) to the ground states. For $\ket{\uparrow \uparrow \downarrow \downarrow}$ the radiated energy will always be lower compared to $\ket{\uparrow \downarrow \uparrow \downarrow}$. We showed that the ground state phase is true ferromagnetic only for $N=2,3$. It is important, as for fermionic ring the ground state is true ferromagnetic only for $N=3$. Usually the superfluid fraction is measured by measuring the energy change due to imposed twisted phase $\Phi_{\nu} \ll \pi$. We showed that apart from already existing methods for generating twisted phase---rotation of either the condensate or the confining potential, and the light atom interaction---one can use the spin generated intrinsic phases $2 \pi p_{\nu}/N_{\nu}$ as a twisted phase. This provides a new way to generate twisted phase in hard-core bosonic rings. We argue that the low lying energy levels occurring due to cyclic permutation of the spin chains ($p_{\nu} \ll N$), can support the persistent current without any external excitation, when: (i) $N \gg 1$, (ii) $N/L \ll 1$. In other words, superfluid emerges spontaneously during transition from fully polarised state ($S=N/2$) to the spin flipped states. In this article we also proposed several experiments to corroborate the above mentioned results.

\begin{acknowledgments}
This article is partly supported by the RFBR Grants No. 21-52-12027. K.K.K would like to thank Pavel D. Grigoriev for useful discussion on the problem.
\end{acknowledgments}

\appendix

\section{Unavailability of $p_{\nu}=0$ state for $\mathrm{S}=(N/2)-1$}
\label{app:pstate}
For the case of $p_{\nu}=0$ the spin wave function $\ket{\psi}$ in Eq. (\ref{eq:spin-eigen-fun}) is symmetric. It corresponds to the fully polarised case. Hence if we have $x$ number of $p_{\nu}=0$ state, then one of the $p_{\nu}=0$ state corresponds to fully polarised case, and $x-1$ number of $p_{\nu}=0$ states corresponds to the spin flipped case. For example, we take a ring with five spinful HCB ($N=5$). The spin configurations for $N=5$ is shown in Tab. \ref{tab:app:spin-config-n5l6}. Here we observe that for $S=5/2$ single $p_{\nu}=0$ state, for $S=3/2$ single $p_{\nu}=0$ state, and for $S=1/2$ two $p_{\nu}=0$ states are available. For $S=3/2$ no $p_{\nu}=0$ is available, because the single $p_{\nu}=0$ corresponds to the fully polarised (ferromagnetic) state. However, for $S=1/2$ because two $p_{\nu}=0$ states are available, one $p_{\nu}=0$ state corresponds to the fully polarised (ferromagnetic) state, and the other $p_{\nu}=0$ corresponds to the non-ferromagnetic phase.

\section{Spin configurations for $N=4$ and $N=5$ \emph{spinful HCB} on a ring}
\label{sec:app:spin-configurations}
In Tab. \ref{tab:app:spin-config-n4l6} we represent all possible spin configurations of four spinful HCB. It should be noted that number of sites $L$ does not have any effect on the spin configurations.
\begin{table}
  \begin{ruledtabular}
    \begin{tabular}{c c c}
      $S$ &$\nu$ &$N_{\nu}$\\
      \hline
      2    &1     &$N_{1}=1$\\
          &      &$p_{1}=0 \equiv \ket{\uparrow \uparrow \uparrow \uparrow}$\\
      \hline
      1    &2     &$N_{2}=4$\\
          &     &$p_{2}=0 \equiv \ket{ \uparrow \uparrow \uparrow \downarrow}$\\
          &     &$p_{2}=1 \equiv \ket{ \downarrow \uparrow \uparrow \uparrow}$\\
          &     &$p_{2}=2 \equiv \ket{ \uparrow \downarrow \uparrow \uparrow}$\\
          &     &$p_{2}=3 \equiv \ket{ \uparrow \uparrow \downarrow \uparrow}$\\
      \hline
      0    &3     &$N_{3}=4$\\
          &      &$p_{3}=0 \equiv \ket{\uparrow \uparrow \downarrow \downarrow}$\\
          &      &$p_{3}=1 \equiv \ket{ \downarrow \uparrow \uparrow \downarrow}$\\
          &      &$p_{3}=2 \equiv \ket{ \downarrow \downarrow \uparrow \uparrow}$\\
          &      &$p_{3}=3 \equiv \ket{ \uparrow \downarrow \downarrow \uparrow}$\\
          &4     &$N_{4}=2$\\
          &      &$p_{4}=0 \equiv \ket{\uparrow  \downarrow \uparrow \downarrow}$\\
          &      &$p_{4}=1 \equiv \ket{ \downarrow \uparrow  \downarrow \uparrow}$\\
    \end{tabular}
  \end{ruledtabular}
\caption{\label{tab:app:spin-config-n4l6} The spin configurations for $N=4$ bosons. Column <<S>> reprsents the total spin of the chain. Column <<$\nu$> represents the enumerated spin blocks. $N_{\nu}$ reprsents the total number of connected spin chains contained in the $\nu$-th spin block. $p_{\nu}$ enumerates the connected spin chains in the $\nu$-th block.}
\end{table}
In Tab. \ref{tab:app:spin-config-n5l6} we represent all possible spin configurations of five spinful HCB.
\begin{table}
  \begin{ruledtabular}
    \begin{tabular}{c c c}
      $S$ &$\nu$ &$N_{\nu}$\\
      \hline
      $\frac{5}{2}$    &1    &$N_{1}=1$\\
          &     &$p_{1}=0 \equiv \ket{\uparrow \uparrow \uparrow \uparrow \uparrow}$\\
      \hline
      $\frac{3}{2}$    &2    &$N_{2}=5$\\
          &     &$p_{2}=0 \equiv \ket{\uparrow \uparrow \uparrow \uparrow \downarrow}$\\
          &     &$p_{2}=1 \equiv \ket{\downarrow \uparrow \uparrow \uparrow \uparrow}$\\
          &     &$p_{2}=2 \equiv \ket{\uparrow \downarrow \uparrow \uparrow \uparrow}$\\
          &     &$p_{2}=3 \equiv \ket{\uparrow \uparrow \downarrow \uparrow \uparrow}$\\
          &     &$p_{2}=4 \equiv \ket{\uparrow \uparrow \uparrow \downarrow \uparrow}$\\
      \hline
      $\frac{1}{2}$    &3    &$N_{3}=5$\\
          &     &$p_{3}=0 \equiv \ket{\uparrow \uparrow \uparrow \downarrow \downarrow}$\\
          &     &$p_{3}=1 \equiv \ket{\downarrow \uparrow \uparrow \uparrow \downarrow}$\\
          &     &$p_{3}=2 \equiv \ket{\downarrow \downarrow \uparrow \uparrow \uparrow}$\\
          &     &$p_{3}=3 \equiv \ket{\uparrow \downarrow \downarrow \uparrow \uparrow}$\\
          &     &$p_{3}=4 \equiv \ket{\uparrow \uparrow \downarrow \downarrow \uparrow}$\\
          &4    &$N_{4}=5$\\
          &     &$p_{4}=0 \equiv \ket{\uparrow \uparrow \downarrow \uparrow \downarrow}$\\
          &     &$p_{4}=1 \equiv \ket{\downarrow \uparrow \uparrow \downarrow \uparrow}$\\
          &     &$p_{4}=2 \equiv \ket{\uparrow \downarrow \uparrow \uparrow \downarrow}$\\
          &     &$p_{4}=3 \equiv \ket{\downarrow \uparrow \downarrow \uparrow \uparrow}$\\
          &     &$p_{4}=4 \equiv \ket{\uparrow \downarrow \uparrow \downarrow \uparrow}$\\
    \end{tabular}
  \end{ruledtabular}
\caption{\label{tab:app:spin-config-n5l6} The spin configurations for $N=5$ bosons. Column <<S>> reprsents the total spin of the chain. Column <<$\nu$> represents the enumerated spin blocks. $N_{\nu}$ reprsents the total number of connected spin chains contained in the $\nu$-th spin block. $p_{\nu}$ enumerates the connected spin chains in the $\nu$-th block.}
\end{table}

\section{Matrix representation of spin wave function}
\label{sec:matr-repr-spin}
For a spin chain configuration $\ket{\uparrow \bullet \uparrow \downarrow}$, using Eq. \eqref{eq:spin-eigen-fun}, the wave functions of the spin chain can be represented in a compact formu sing matrix notation $\ket{\psi(p_{\nu})} = (1/\sqrt{N_{\nu}}) \mathrm{C} \ket{\tilde{\psi}_{\nu}}$. Here $\ket{\psi(p_{\nu})}$ is a $N_{\nu} \times 1$ column matrix. Its components represent the wave function corresponding to $p_{\nu}$-th value. $\mathrm{C}$ is the $N_{\nu} \times N_{\nu}$ matrix. Its rows and columns are indexed as $p_{\nu} = 0, 1, \dots N_{\nu}$. Hence $\mathrm{C}_{mn}$-th terms is $\mathrm{e}^{\imath 2 \pi m (n/N_{\nu})}$. The wave function $\ket{\tilde{\psi}_{\nu}}$ is a $N_{\nu} \times 1$ column matrix of all possible connected spin chain of the $\nu$-th block. If the $\nu$-th block represents all the connected spin chain of $\ket{\uparrow \bullet \uparrow \downarrow}$, the total wave function of the $\nu$-th block is:
\begin{equation}
  \label{eq:app:spin-wave-mat}
    \begin{bmatrix}
      \psi_{\nu}(p_{\nu}=0)\\
      \psi_{\nu}(p_{\nu}=1)\\
      \psi_{\nu}(p_{\nu}=2)
    \end{bmatrix}
    = \frac{1}{\sqrt{N_{\nu}}}
    \begin{bmatrix}
      1 &1 &1 \\
      1 &\mathrm{e}^{\imath 2 \pi/3} &\mathrm{e}^{\imath 4 \pi/3}\\
      1 &\mathrm{e}^{\imath 4 \pi/3} &\mathrm{e}^{\imath 8 \pi/3}
    \end{bmatrix}
    \begin{bmatrix}
      \ket{\uparrow \bullet \uparrow \downarrow}\\
      \ket{\downarrow \bullet \uparrow \uparrow}\\
      \ket{ \uparrow  \bullet \downarrow \uparrow}
    \end{bmatrix}
    .
  \end{equation}
Eq. (\ref{eq:app:spin-wave-mat}) can be generalized to arbitrary $N$.

\end{document}


\title{Suplementary Materials to: Spin-flip induced superfluidity in a ring of spinful hard-core bosons}
\begin{abstract}
  In Sec. \ref{sec:deta-deriv-energy} we gave a detailed derivation of the energy levels in the presence and absence of magnetic field, for periodic and anti-periodic boundary condition. In Sec. \ref{sec:app:repr-hcb-hamilt} we pedagogically discuss the transformation of bosonic Hamiltonian into fermionic Hamiltonian using Jordan-Wigner transformation. In Sec. \ref{app:groundenergy} ground state energy for every one of the eight cases has been calculated.

\end{abstract}
\maketitle

\section{A detailed derivation of the energy levels}
\label{sec:deta-deriv-energy}

\subsection{The Hamiltonian and energy levels without magnetic field}
\label{sec:hamilt-spinf-hcb}

We take a ring of $L$ sites and $N$ spinful HCB. In the ring every site contains \emph{at most single} boson ($N \leq L$) because of the hard-core nature of the bosons. Although, the bosons don't have two spin projections $\pm 1/2$ like fermions, but two \emph{hyperfine} states of some atoms can be considered as two spin projections of the bosons. These spins are known as the \emph{pseudo spins}, and thes atoms are known as the \emph{spinor bosons} \cite{stamper-kurn-SpinorBoseGases-2013,kawaguchi-SpinorBoseEinstein-2012}. Naturally, one can map the \emph{pseudo spins} to the spin-up state $(\ket{\uparrow})$ and the spin-down state $(\ket{\downarrow})$. We start with the basic \emph{t-J$_{z}$} Hamiltonian \footnote{The \emph{t-J$_{z}$} Hamiltonian is the usual \emph{t-J} model in strong on-site repulsive potential}:
\begin{equation}
  \label{eq:t-J-ham}
  H = - \sum\limits_{i j \sigma} t_{ij} \tilde{b}_{i \sigma}^{\dagger} \tilde{b}_{j \sigma} + J \sum\limits_{ij} \left(S_{iz}S_{jz}\right). 
\end{equation}
Here, $t_{ij}$ is the boson hopping factor from the \emph{i}-th site to the \emph{j}-th site. $\tilde{b}_{i \sigma}^{\dagger} \equiv \left( 1-n_{i, -\sigma} \right) b_{i \sigma}^{\dagger}$ and $\tilde{b}_{i \sigma} \equiv b_{i \sigma} \left( 1-n_{i, -\sigma} \right)$ are the \emph{constraint} HCB creation and annihilation operators with spin projections $\sigma=\left\{ \uparrow, \downarrow \right\}$. $n_{i, \sigma} \equiv b_{i \sigma}^{\dagger} b_{i \sigma}$ is the boson number operator. The operators $\tilde{b}_{i \sigma}$ and $\tilde{b}_{i \sigma}^{\dagger}$ implicitly contain the single occupancy constraint. $J = (2 t^{2}/U)$ is the indirect exchange integral; $U$ is the on-site repulsive potential. The
\begin{equation}
  \label{eq:spin-opr}
   S_{iz}= (1/2) \left( \tilde{b}_{i,\uparrow}^{\dagger} \tilde{b}_{i,\uparrow} - \tilde{b}_{i,\downarrow}^{\dagger} \tilde{b}_{i,\downarrow}  \right)
\end{equation}
is the spin operator in terms of the constraint boson HCB operators. For infinite potential $(U/t \approx \infty)$, known as the \emph{Nagaoka limit}, the second term in the Hamiltonian vanishes $(J=0)$ \cite{nagaoka-FerromagnetismNarrowAlmost-1966}. Hence, we are left with the simplified Hamiltonian:
\begin{equation}
  \label{eq:t-J-Ham-simp}
  H = - \sum\limits_{i j \sigma} t_{ij} \tilde{b}_{i \sigma}^{\dagger} \tilde{b}_{j \sigma}.
\end{equation}

We assume that, the boson hopping is isotropic and homogeneous. Additionally, we allow hopping only to the nearest neighbours. Hence, in the $t_{ij}$ matrix only the $t_{i \pm 1}=t$ terms are non zero; while all other terms are zero. A 1D chain is equivalent to the ring, if bosons hop directly from the \emph{L}-th site to the \emph{1}-st site. Two types of boundary conditions are possible for a ring: (i) the periodic boundary condition ($\tilde{b}_{L+1, \sigma} = \tilde{b}_{1, \sigma}$), (ii) the antiperiodic boundary condition ($\tilde{b}_{L+1, \sigma} = -\tilde{b}_{1, \sigma}$). Although, the Hamiltonian and the energy levels for both these cases differ slightly, but all calculations are analogously identical. Hence, we will show all calculations for the periodic boundary condition, and will mention only the important results for antiperiodic boundary condition. Therefore, using the aforementioned conditions, the Hamiltonian for spinful HCB on a ring in the \emph{Nagaoka limit} and with the periodic boundary condition is:
\begin{equation}
  \label{eq:Ham-with-spin}
  H = - t \sum\limits_{i=1, \sigma}^{L-1} \tilde{b}_{i \sigma}^{\dagger} \tilde{b}_{i+1 \sigma} - t \sum\limits_{\sigma} \tilde{b}_{L \sigma}^{\dagger} \tilde{b}_{1 \sigma} + \mathrm{H.c}.
\end{equation}
The second term in Eq. (\ref{eq:Ham-with-spin}) represents the jump of the bosons from the $L$-th site to the \emph{1}-st site.

In Eq. (\ref{eq:Ham-with-spin}) we can separate the spin and the spatial part of the \emph{N}-bosonic wave function because of the absence of the spin interactions. If $\ket{\psi}$ is the \emph{N}-bosonic wave function, then $\ket{\psi} = \ket{\psi_{r}} \ket{\psi_{s}}$. Here, $\ket{\psi_{r}} = \ket{r_{1}, r_{2}, \dots, r_{N}}$ is the spatial part, and $\ket{\psi_{s}} = \ket{s_{1}^{z}, s_{2}^{z}, \dots, s^{z}_{N}}$ is the spin part. In the ring the bulk hopping of the bosons results in the cyclic permutation of the spins, because the single occupancy constraint should be satisfied. Therefore, we introduce the spin permutation operator $\hat{P}$. It operates only on the spin of the particles. It displaces the spin to the next non-empty site:
\begin{equation}
  \label{eq:spin-permut}
  \hat{P} \ket{s_{1}^{z}, s_{2}^{z}, \dots, s^{z}_{N}} = \ket{s_{N}^{z}, s_{1}^{z}, \dots, s^{z}_{N-1}}.
\end{equation}
Using $\hat{P}$, we write Eq. (\ref{eq:Ham-with-spin}) as:
\begin{equation}
  \label{eq:Ham-hcb}
  H = -t \sum\limits_{i=1}^{L-1} \tilde{b}_{i}^{\dagger} \tilde{b}_{i+1} -t \tilde{b}_{L}^{\dagger} \tilde{b}_{1} \hat{P} + \mathrm{H.c}.
\end{equation}
Here, $\tilde{b}_{i}^{\dagger} (\tilde{b}_{i})$ is the \emph{spinless} HCB creation (annihilation) operator. On the same site they obey following commutation rules:
\begin{equation}
  \label{eq:rules-commut-same}
  \begin{aligned}
    \left\{ \tilde{b}_i^{\dagger},\tilde{b}_i\right\}= 1, \quad \tilde{b}_{i}^{\dagger}\tilde{b}_{i}^{\dagger} =0, \quad \tilde{b}_i\tilde{b}_i=0 .
  \end{aligned}
\end{equation}
The fermionic nature of these rules is clear. Similarly, on different sites they obey following rules:
\begin{equation}
  \label{eq:nearsite-commutation-hcb}
  \left[\tilde{b}_i^{\dagger}, \tilde{b}_j \right]= \left[ \tilde{b}_{i}^{\dagger},\tilde{b}_{j}^{\dagger} \right] = \left[ \tilde{b}_i, \tilde{b}_j\right]=0, \: \forall \: i \neq j.
\end{equation}
The bosonic nature of these rules is clear. \emph{Hence, spinless HCB behaves like bosons on different sites, and like fermions at the same sites.} Because of this behaviour, the Hamiltonian in Eq. (\ref{eq:Ham-hcb}) can not be diagonalised in the present form. However, using the transformation $\hat{\sigma}_{i}^{+} \equiv \tilde{b}^{\dagger}_{j} \sqrt{1 - n_{j}}$, and $\hat{\sigma}_{i}^{-} \equiv \sqrt{1 - n_{j}} \tilde{b}_{j}$, Eq. (\ref{eq:Ham-hcb}) can be mapped to the \emph{XY} spin-chain Hamiltonian \cite{holstein-FieldDependenceIntrinsic-1940}. Further, one uses \emph{Jordan-Wigner transformation} (JWT) to represent the \emph{XY} spin-chain Hamiltonian in terms of the fermionic operators, which obeys fermionic commutation rules everywhere \cite{cazalilla-OneDimensionalBosons-2011,lieb-TwoSolubleModels-1961,wang-SpontaneousAharonovCasherEffect-1995}. The new Hamiltonian can be diagonalised easily.

The Hamiltonian of Eq. (\ref{eq:Ham-hcb}) in terms of the fermionic operators is [see App. \ref{sec:app:repr-hcb-hamilt}]:
\begin{equation}
  \label{eq:Hamiltonian-ferm-opr-period}
  H = - t \sum\limits_{i=1}^{L-1}\hat{f}_i^{\dagger}\hat{f}_{i+1} + \: t \: \mathrm{e}^{\imath \pi N} \hat{f}_{L}^{\dagger}\hat{f}_1\: \hat{P}+ \text{H.c}.
\end{equation}
Here, $\hat{f}_i^{\dagger}$ ($\hat{f}_i$) is the spinless fermionic creation (annihilation) operator. We absorb the <<+>> sign in front of the second term into the phase to give the Hamiltonian more symmetrical form: $+t e^{\imath \pi N}\hat{f}_{L}^{\dagger} \hat{f}_{1}\hat{P} = -t e^{\imath \pi (N-1)}\hat{f}_{L}^{\dagger} \hat{f}_{1}\hat{P}$. The effect of the phase $\mathrm{e}^{\imath \pi (N-1)}$ is only to change the sign in front of the second term when $N$ is even; when $N$ is odd the sign does not change. Hence, we introduce a two valued function $h(N)$ to take care of this:
\begin{equation}
  \label{eq:Hn-funct}
  h(N) \equiv
  \begin{cases}
    0, \quad \text{odd }N;\\
    1, \quad \text{even }N.
  \end{cases}
\end{equation}
Substituting $h(N)$ into Eq. \eqref{eq:Hamiltonian-ferm-opr-period}, we get:
\begin{equation}
  \label{eq:Ham-ferm-period-odd-even}
    H = - t \sum\limits_{i=1}^{L-1}\hat{f}_i^{\dagger}\hat{f}_{i+1} - \: t \: \mathrm{e}^{\imath \pi h(N)} \hat{f}_{L}^{\dagger}\hat{f}_1\: \hat{P}+ \text{H.c}.
\end{equation}
It should be noted that, for odd $N$ the Hamiltonian in Eq. (\ref{eq:Ham-ferm-period-odd-even}) is exactly equal to the Hamiltonian for the fermionic case \cite{ivantsov-StableMetastableKinetic-2020}.

\emph{Therefore, a HCB ring with periodic boundary condition behaves as a fermionic ring when the number of particles in the ring is odd. Similarly, one can easily deduce that, a HCB ring with antiperiodic boundary condition behaves as a fermionic ring when the number of particles in the ring is even.}

The energy levels of the HCB ring are found by diagonalising the Hamiltonian. In Eq. \eqref{eq:Ham-ferm-period-odd-even} we diagonalise the spin permutation operator and fermionic operators separately because they are independent of each other. The eigen values and eigen function of $\hat{P}$ are same as the momentum ($\lambda_{\nu}$) and the eigen function ($\psi_{\nu}$) of the spin waves \cite{ivantsov-StableMetastableKinetic-2020}:
\begin{subequations}
\begin{equation}
  \label{eq:spin-eigen-value}
  \begin{aligned}
    &\lambda_{\nu} = \mathrm{e}^{\imath 2 \pi p_{\nu}/ N_{\nu}};\\
  \end{aligned}
\end{equation}    
\begin{equation}
  \label{eq:spin-eigen-fun}
  \begin{aligned}    
    &\ket{\psi_{\nu}} = \frac{1}{\sqrt{N_{\nu}}} \sum\limits_{q=0}^{N_{\nu}-1} \mathrm{e}^{\imath 2 \pi p_{\nu} \frac{q}{N_{\nu}}} \hat{P}^q \: \ket{ \tilde{\psi}_{\nu}}.
  \end{aligned}
\end{equation}
\end{subequations}
Here, $\nu$ enumerates all the possible disconnected spin blocks. A spin block contains only connected spin chains. When two spin chains can be transformed to each other by application of $\hat{P}$ operator they are connected, otherwise they are disconnected. $N_{\nu}$ represents the total number of connected spin chains in $\nu$-th spin block. $p_{\nu}$ enumerates the connected spin chains in the $\nu$-th spin block; it takes the values $p_{\nu}= 0, 1, \dots, N_{\nu}-1$. $\tilde{\psi}_{\nu}$ is the wave function of one of the spin chain of $\nu$-th spin block. It should be noted that, the $p_{\nu}=0$ state is not available for all spin blocks.

For example, we have a chain of 4 sites and 3 particles. We take the spin-chain $\ket{\uparrow \bullet \uparrow \downarrow}$ out of $2^{3}$ possible spin-chains. It is connected to the $\ket{\uparrow \bullet \downarrow \uparrow}$ spin-chain, as $\hat{P}^{2} \ket{\uparrow \bullet \uparrow \downarrow} = \ket{\uparrow \bullet \downarrow \uparrow}$. In this case both these configurations belong to the same $\nu$-th spin block. This particular $\nu$-th block will have 3 possible configurations, hence $N_{\nu}$ = 3, and $p_{\nu} = 0, 1, 2$ \footnote{The three configurations ($N_{\nu}$ = 3) corresponding to three $p_{\nu}$ are: (i) $p_{\nu}=0 \equiv \ket{\uparrow \bullet \uparrow \downarrow}$, (ii) $p_{\nu}=1 \equiv \ket{\downarrow \bullet \uparrow \uparrow }$, (iii) $p_{\nu}=2 \equiv \ket{\uparrow \bullet \downarrow \uparrow}$.}. The wave function of three $p_{\nu}$ states can be found using Eq. (\ref{eq:spin-eigen-fun}). Consequently, every spin-chain in the $\nu$-th spin block has its own wave function and spin momentum $p_{\nu}$. The number of disconnected blocks will depend on the number of particles present in the ring ($N$) and the spin of these particles ($s^{z}$). Due to these disconnected block of spin chains, the total spin Hamiltonian corresponding to the $\hat{P}$ operator is a block Hamiltonian with $\nu$ blocks. We find the Hamiltonian corresponding to the $\nu$-th block by substituting $\lambda_{\nu}$ from Eq. (\ref{eq:spin-eigen-fun}) into Eq. (\ref{eq:Ham-ferm-period-odd-even}):
\begin{equation}
  \label{eq:ham-spin-opr}
  H_{\nu} =
    - t \sum\limits_{i=1}^{L-1}\hat{f}_i^{\dagger}\hat{f}_{i+1} - \: t \: \mathrm{e}^{\imath 2 \pi \left( \frac{p_{\nu}}{N_{\nu}} + \frac{h(N)}{2}\right)} \hat{f}_{L}^{\dagger}\hat{f}_1 + \text{H.c}.
\end{equation}
The total Hamiltonian of the whole system is found as tensor product of these spin block Hamiltonians: $H = \sum\limits_{\nu} \otimes H_{\nu}$.

Eq. \eqref{eq:ham-spin-opr} is nothing but the tight binding model with a penetrating magnetic flux $\Phi_{\nu} \equiv 2 \pi \left[ \frac{p_{\nu}}{N_{\nu}} + \frac{h(N)}{2}\right]$ through the ring. One makes the gauge transformation $f_{i} \mapsto \mathrm{e}^{\imath \Phi_{\nu} x_{i}/L} f_{i}$ to solve this Hamiltoinian. Here, $x_{i} = 1, 2, \dots, L$, enumerates the $L$ sites. This process is known as the \emph{Peierls substitution}. Eq. (\ref{eq:ham-spin-opr}) after the gauge transformation becomes:
\begin{equation}
  \label{eq:ham-gauge}
  H_{\nu}=
      - t \sum\limits_{i}^{L-1}\mathrm{e}^{\imath \frac{\Phi_{\nu}}{L}}\hat{f}_i^{\dagger}\hat{f}_{i+1}
      - \: t \: \mathrm{e}^{\imath \frac{\Phi_{\nu}}{L}} \hat{f}_{L}^{\dagger}\hat{f}_1 + \text{H.c}.
\end{equation}
The $k$-th mode energy of this tight binding Hamiltoinian $H_{\nu}$ is:
\begin{subequations}
\begin{equation}
  \label{eq:eigen-energy-even}
  \begin{aligned}
  E_{\text{PBC}}&(k, \nu, p_{\nu}; N, L)=\\
  &-2t \cos \frac{2 \pi}{L}\left( k+ \frac{p_{\nu}}{N_{\nu}} + \frac{1}{2} \right),\:&& \text{even } N;\\
  \end{aligned}  
\end{equation}  
\begin{equation}
  \label{eq:eigen-energy-odd}
  \begin{aligned}
  E_{\text{PBC}}&(k, \nu, p_{\nu}; N, L)=\\  
    &-2t \cos \frac{2 \pi}{L}\left( k+ \frac{p_{\nu}}{N_{\nu}} \right),\:&& \text{odd } N.
  \end{aligned}  
\end{equation}
\end{subequations}
Here, $k$ is the wave vector. It is quanatized and takes the values $k=0, \pm 1, \pm 2, \dots$. The $k$-th mode energy for antiperiodic boundary condition is analogous to Eq. \eqref{eq:eigen-energy-odd} and \eqref{eq:eigen-energy-even}:
\begin{subequations}
\begin{equation}
  \label{eq:eigen-energy-apbc-even}
  \begin{aligned}
  E_{\text{APBC}}&(k, \nu, p_{\nu}; N, L)=\\
  &-2t \cos \frac{2 \pi}{L}\left( k+ \frac{p_{\nu}}{N_{\nu}} \right),\:&& \text{even } N;\\
  \end{aligned}  
\end{equation}  
\begin{equation}
  \label{eq:eigen-energy-apbc-odd}
  \begin{aligned}
  E_{\text{APBC}}&(k, \nu, p_{\nu}; N, L)=\\    
    &-2t \cos \frac{2 \pi}{L}\left( k+ \frac{p_{\nu}}{N_{\nu}} + \frac{1}{2}\right),\:&& \text{odd } N.
  \end{aligned}  
\end{equation}
\end{subequations}
\emph{Comparing Eq. \eqref{eq:eigen-energy-apbc-even} and \eqref{eq:eigen-energy-odd}, we note that, the k-th mode energy for periodic boundary condition and odd $N$ is equal to the k-th mode energy of antiperiodic boundary condition and even $N$. Analogously, the k-th mode energy for periodic boundary condition and even $N$ is equal to the k-th mode energy for antiperiodic boundary condition and odd $N$.}

\subsection{The Hamiltonian and energy levels with magnetic field}
\label{sec:hamilt-energy-levels-with-mag}
It is interesting to investigate the behaviour of HCB ring under the magnetic field. As before, the Hamiltonian can be solved for both periodic and antiperiodic boundary condition. However, we will show the calculations only for the periodic boundary condition. The calculations for the antiperiodic boundary condition can be done analogously. We assume that, an uniform and constant magnetic field is appied perpendicular to the plane of the ring. The Hamiltonian in Eq. \eqref{eq:ham-gauge} will have two additional terms because of the presence of magnetic field: (i) additional phase contribution due to the \emph{Aharonov-Bohm effect} \cite{aharonov-SignificanceElectromagneticPotentials-1959}, (ii) additional energy contribution due to the \emph{Zeeman effect} \cite{condon-TheoryAtomicSpectra-1935}. The Hamiltonian is:
\begin{equation}
  \label{eq:ham-mag-field}
  \begin{aligned}
  H_{\nu}=
      - t &\sum\limits_{i}^{L-1}\mathrm{e}^{\imath \frac{2\pi}{L} \left( \frac{p_{\nu}}{N_{\nu}} + \frac{h(N)}{2} + \frac{\Phi_{B}}{\Phi_0} \right)}\hat{f}_i^{\dagger}\hat{f}_{i+1}\\
      - &\: t \: \mathrm{e}^{\imath \frac{2\pi}{L} \left( \frac{p_{\nu}}{N_{\nu}} + \frac{h(N)}{2} + \frac{\Phi_{B}}{\Phi_0}\right)} \hat{f}_{L}^{\dagger}\hat{f}_1 + \text{H.c.}\\ &+ g \mu_{B} B S_{\nu z}.
  \end{aligned}    
\end{equation}
Here, $\Phi_{B}$  is the total magnetic flux penetrating through the ring, $\Phi_{0}=h/e$ is the \emph{magnetic flux quanta}, $g$ is the \emph{Lande factor}, $\mu_{B}$ is the \emph{Bohr magneton}, $B$ is the applied magnetic flux density, and $S_{\nu z}$ is the total spin projection along z-axis of the $\nu$-th spin-block. It is found by summing the spin projections along \emph{z}-axis of $1,2,\dots,L$-th sites. The last term in Eq. \eqref{eq:ham-mag-field} is known as the \emph{Zeeman energy}. It just adds a constant and equal energy to all the energy levels corresponding to the spin chains ($p_{\nu}$) of the $\nu$-th spin block. It depends on the applied magnetic flux density $B$, and the total spin of the $\nu$-th spin block. We can define the \emph{Zeeman energy} of the $\nu$-th spin block as:
\begin{equation}
  \label{eq:zeeman-ener}
  Z_{\nu} \equiv g \mu_{B} B S_{\nu z}.
\end{equation}
$\Phi_{B}$ is found by multiplying the area ($A$) of the ring with magnetic flux density: $\Phi_{B}=B \cdot A$. Eq. (\ref{eq:ham-mag-field}) is analogous to Eq. (\ref{eq:ham-gauge}), apart from the last term. Therefore, the $k$-th mode energy of the Hamiltonian under magnetic field will be analogous to Eq. (\ref{eq:eigen-energy-even}) and (\ref{eq:eigen-energy-odd}). The $k$-th mode energy is:
\begin{subequations}
\begin{equation}
  \label{eq:eigen-energy-b-field-even}
  \begin{aligned}
    &E^{(B)}_{\text{PBC}}(k, \nu, p_{\nu}; N, L)=\\
    &-2t \cos \frac{2 \pi}{L}\left( k+ \frac{p_{\nu}}{N_{\nu}} + \frac{1}{2} + \frac{\Phi_{B}}{\Phi_0} \right) + Z_{\nu},\: &&\text{even } N;
    \end{aligned}
  \end{equation}
\begin{equation}
  \label{eq:eigen-energy-b-field-odd}
  \begin{aligned}
    &E^{(B)}_{\text{PBC}}(k, \nu, p_{\nu}; N, L)=\\
     &-2t \cos \frac{2 \pi}{L}\left( k+ \frac{p_{\nu}}{N_{\nu}} + \frac{\Phi_{B}}{\Phi_0} \right) + Z_{\nu},\: &&\text{odd } N.
  \end{aligned}
\end{equation}
\end{subequations}
Analogously, for the antiperiodic boundary condition the $k$-th mode energy is:
\begin{subequations}
\begin{equation}
  \begin{aligned}
  \label{eq:eigen-energy-b-field-apbc-even}
  &E^{(B)}_{\text{APBC}}(k, \nu, p_{\nu}; N, L)=\\
  &-2t \cos \frac{2 \pi}{L}\left( k+ \frac{p_{\nu}}{N_{\nu}}  + \frac{\Phi_{B}}{\Phi_0} \right) + Z_{\nu},  &&\text{even } N;\\
  \end{aligned}
\end{equation}
\begin{equation}
  \begin{aligned}
    \label{eq:eigen-energy-b-field-apbc-odd}
  &E^{(B)}_{\text{APBC}}(k, \nu, p_{\nu}; N, L)=\\    
    &-2t \cos \frac{2 \pi}{L}\left( k+ \frac{p_{\nu}}{N_{\nu}} + \frac{1}{2} + \frac{\Phi_{B}}{\Phi_0} \right) + Z_{\nu}, &&\text{odd } N.
\end{aligned}
\end{equation}
\end{subequations}
\emph{Comparison of above equations shows that, the k-th mode energy for periodic boundary condition and odd N is equal to the k-th mode energy for antiperiodic boundary condition and even N. Similarly, the k-th mode energy for periodic boundary condition and even N is equal to the k-th mode energy for antiperiodic boundary condition and odd N.}

\section{Representation of HCB Hamiltonian in terms of fermionic operator}
\label{sec:app:repr-hcb-hamilt}

In this section we show the transformation of the spinful HCB Hamiltonian on a ring from bosonic operator into the fermionic operator. The transformation process has already been outlined in Ref. \cite{cazalilla-OneDimensionalBosons-2011}. However, we show here the complete calculation. We start from the Hamiltonian in Eq. (\ref{eq:Ham-hcb}):
\begin{equation}
  \label{eq:app:Ham-hcb}
  H = -t \sum\limits_{i=1}^{L-1} \tilde{b}_{i}^{\dagger} \tilde{b}_{i+1} -t \tilde{b}_{L}^{\dagger} \tilde{b}_{1} \hat{P} + \mathrm{H.c}.
\end{equation}
We use the following transformations to represent the above Hamiltonian in terms of spin operators \cite{holstein-FieldDependenceIntrinsic-1940}:
\begin{equation}
  \label{eq:app:holst-trans}
  \hat{\sigma}_{i}^{+} = \tilde{b}_{i}^{\dagger} \sqrt{1- \hat{n}_{i}};\quad \hat{\sigma}_{i}^{-} = \sqrt{1- \hat{n}_{i}} \tilde{b}_{i}.
\end{equation}
Here $\hat{\sigma}^{+}$ and $\hat{\sigma}^{-}$ are the spin raising and lowering operators. These are defined as:
\begin{equation}
  \label{eq:app:rais-low-opr}
  \hat{\sigma}_{i}^{+} = \hat{\sigma}_{i}^{x} + \imath \sigma_{i}^{y}; \quad \hat{\sigma}_{i}^{-} = \hat{\sigma}_{i}^{x} - \imath \sigma_{i}^{y}
\end{equation}
Here $\hat{\sigma}_{i}^{x}$ and $\hat{\sigma}_{i}^{y}$ are the spin projections along \emph{x} and \emph{y} axes. Using Eq. (\ref{eq:app:rais-low-opr}) we write the Hamiltonian as:
\begin{equation}
  \label{eq:app:Ham-hcb-sigma}
  H = -t \sum\limits_{i=1}^{L-1} \hat{\sigma}_{i}^{+} \hat{\sigma}_{i+1}^{-} -t \hat{\sigma}_{L}^{+} \hat{\sigma}_{1}^{-} \hat{P} + \mathrm{H.c}.
\end{equation}
Substituting Eq. (\ref{eq:app:rais-low-opr}) into Eq. (\ref{eq:app:Ham-hcb-sigma}) we find:
\begin{equation}
  \label{eq:app:Ham-hcb-XY}
  \begin{aligned}
    H = -2t \sum\limits_{i=1}^{L-1}& \left(\hat{\sigma}_{i}^{x} \hat{\sigma}_{i+1}^{x} + \hat{\sigma}_{i}^{y} \hat{\sigma}_{i+1}^{y} \right)\\ &-2t \left( \hat{\sigma}_{L}^{x} \hat{\sigma}_{1}^{x} + \hat{\sigma}_{L}^{y} \hat{\sigma}_{1}^{y} \right) \hat{P} + \mathrm{H.c}.
    \end{aligned}
\end{equation}
Eq. (\ref{eq:app:Ham-hcb-sigma}) is analogous to \emph{XY} spin-chain Hamiltonian. One can use Jordan-Wigner transformation (\emph{JWT}) to solve the \emph{XY} spin-chain model \cite{lieb-TwoSolubleModels-1961}. We use the \emph{JWT} to represent the Hamiltonian in Eq. (\ref{eq:app:Ham-hcb-sigma}) in terms of fermionic operators \cite{lieb-TwoSolubleModels-1961}.
In JWT the spin raising and lowering operators are represented as \cite{lieb-TwoSolubleModels-1961}:
\begin{equation}
  \label{eq:app:jwt-tranformation}
  \begin{aligned}
    \hat{\sigma}_i^{+} = \hat{f}_i^{\dagger} \mathrm{e}^{-\imath \hat{\phi}_{i}}; \:
    \hat{\sigma}_i^{-} =  \mathrm{e}^{\imath \hat{\phi}_{i}} \hat{f}_i.
  \end{aligned}
\end{equation}
Here $\hat{f}_i^{\dagger}$ ($\hat{f}_i$) is the spinless fermionic creation (annihilation) operator. $\hat{\phi}_{i}$ is the string operator defined as:
\begin{equation}
  \label{eq:app:string-operator}
  \hat{\phi}_i = \pi \sum\limits_{i=1}^{i-1} \hat{n}_{i}.
\end{equation}
Here $\hat{n}_{i} = \hat{f}_i^{\dagger} \hat{f}_i$ is the occupation number operator. In Eq. (\ref{eq:app:string-operator}) the summation is taken over all the sites present to the left of the $i$-th site. Hence for every non empty site a phase factor $\pi$ is multiplied to the fermionic operator in Eq. (\ref{eq:app:jwt-tranformation}). Substituting Eq. (\ref{eq:app:jwt-tranformation}) in Eq. (\ref{eq:app:Ham-hcb-sigma}) the Hamiltonian can be represented in spinless fermionic operators. The first term of the Hamiltonian is:
\begin{equation}
  \label{eq:1st-term-ham-ferm-opr}
  \begin{aligned}
    \hat{\sigma}_j^{+} \hat{\sigma}_{j+1}^{-} &= \hat{f}_j^{\dagger} \mathrm{e}^{-\imath \hat{\phi}_{j}} \mathrm{e}^{\imath \hat{\phi}_{j+1}} \hat{f}_{j+1} =  \hat{f}_j^{\dagger} \mathrm{e}^{-\imath \pi \hat{n}_j} \hat{f}_{j+1} = \hat{f}_j^{\dagger} \hat{f}_{j+1}.
  \end{aligned}  
\end{equation}
In Eq. (\ref{eq:1st-term-ham-ferm-opr}) as creation operator $\hat{f}_j^{\dagger}$ is present to the left of $\mathrm{e}^{-\imath \pi n_{j}}$, the $j$-th site should be empty ($n_j=0$). Hence, the exponential term $\mathrm{e}^{-\imath \pi n_{j}}$ will not have any effect.

The second term is more tricky. Due to periodic boundary condition, we can write the $\hat{\sigma}_{1}^{-}=\hat{\sigma}_{L+1}^{-}$. Substituting Eq. (\ref{eq:app:jwt-tranformation}) in the second term of Eq. (\ref{eq:app:Ham-hcb-sigma}), and using the condition $\hat{\sigma}_{1}^{-}=\hat{\sigma}_{L+1}^{-}$ we find:
\begin{equation}
  \label{eq:app:ham-second-term}
  t \hat{\sigma}^{+}_{L} \hat{\sigma}^{-}_{1} \hat{P} = t \hat{\sigma}^{+}_{L} \hat{\sigma}^{-}_{L+1} \hat{P} = t \hat{f}^{\dagger}_{L} \mathrm{e}^{\imath \pi \hat{n}_{L}} \hat{f}_{L+1} \hat{P} = t \hat{f}^{\dagger}_{L}  \hat{f}_{L+1} \hat{P}.
\end{equation}
Now we will find $\hat{f}_{L+1}$ in terms of $\hat{f}_{1}$. By definition of the periodic boundary condition we have:
  \begin{equation}
    \label{eq:app:ferm-opr-sec-term}
    \hat{f}_{L+1}\psi_{F}(x_{2},x_{3},\dots,x_{L}, x_{L+1}) = \hat{f}_{1}\psi_{F}(x_{1},x_{2},\dots,x_{L}).
  \end{equation}
Here, $\psi_{F}$ is the fermionic asymmetric wave function. $x_{1},x_{2}$ are the site numbers on a 1D chain. Due to periodic boundary condition we can write $\psi_{F}(x_{1},x_{2},\dots,x_{L}) = \psi_{F}(x_{L+1},x_{2},\dots,x_{L})$. Hence the right hand side of Eq. (\ref{eq:app:ferm-opr-sec-term}) can be written as:
\begin{equation}
    \label{eq:app:ferm-opr-sec-term-2}
\hat{f}_{1}\psi_{F}(x_{1},x_{2},\dots,x_{L}) = \hat{f}_{1}\psi_{F}(x_{L+1},x_{2},\dots,x_{L}).
  \end{equation}
As $\psi_{F}$ is the asymmetric fermionic wave function we can write:
\begin{equation}
  \label{eq:app:ferm-opr-sec-term-3}
  \begin{aligned}
    \psi_{F}(x_{L+1},x_{2},\dots,x_{L}) &= (-1)^{L-1} \psi_{F}(x_{2},\dots,x_{L},x_{L+1})\\
    &= \mathrm{e}^{\imath \pi (L-1)} \psi_{F}(x_{2},\dots,x_{L},x_{L+1}). 
  \end{aligned}
  \end{equation}  
Substituting the value of $\psi_{F}(x_{L+1},x_{2},\dots,x_{L})$ from Eq. (\ref{eq:app:ferm-opr-sec-term-3}) on the right hand side of Eq. (\ref{eq:app:ferm-opr-sec-term-2}) we will get:
\begin{equation}
    \label{eq:app:ferm-opr-sec-term-4}
\hat{f}_{1}\psi_{F}(x_{1},x_{2},\dots,x_{L}) = \hat{f}_{1}\mathrm{e}^{\imath \pi (L-1)} \psi_{F}(x_{2},\dots,x_{L},x_{L+1}). 
  \end{equation}
  Substituting Eq. (\ref{eq:app:ferm-opr-sec-term-4}) on the right hand side of Eq. (\ref{eq:app:ferm-opr-sec-term}) we will get:
  \begin{equation}
    \label{eq:app:ferm-opr-sec-term-5}
    \begin{aligned}
         \hat{f}_{L+1}&\psi_{F}(x_{2},x_{3},\dots,x_{L}, x_{L+1}) \\ &=  \hat{f}_{1}\mathrm{e}^{\imath \pi (L-1)} \psi_{F}(x_{2},\dots,x_{L},x_{L+1}).
    \end{aligned}
  \end{equation}  
  If we compare the right and left hand side of Eq. (\ref{eq:app:ferm-opr-sec-term-5}) we will get:
  \begin{equation}
    \label{eq:app:rel-ferm-opr}
    \hat{f}_{L+1} = \hat{f}_{1}\mathrm{e}^{\imath \pi (L-1)} = - \hat{f}_{1} e^{\imath \pi L}.
  \end{equation}
  It should be noted that for our case in Eq. (\ref{eq:app:rel-ferm-opr}) $L$ is equivalent to the number of particles $N$. Substituting value of $\hat{f}_{L+1}$ into Eq. (\ref{eq:app:ham-second-term}) we will find:
\begin{equation}
  \label{eq:app:ham-second-term-2}
  t \hat{\sigma}^{+}_{L} \hat{\sigma}^{-}_{1} \hat{P} = t \hat{f}^{\dagger}_{L}  \hat{f}_{L+1} \hat{P} =  - t \hat{f}^{\dagger}_{L}  \hat{f}_{1} \hat{P} \mathrm{e}^{\imath \pi N}.
\end{equation}  
Using Eq. (\ref{eq:app:ham-second-term-2}) and (\ref{eq:1st-term-ham-ferm-opr}) in Eq. (\ref{eq:app:Ham-hcb-sigma}) we will find the Hamiltonian for periodic boundary condition in terms of the fermionic operators:
\begin{equation}
  \label{eq:app:Ham-ferm}
  H = -t \sum\limits_{i=1}^{L-1} \hat{f}_{i}^{\dagger} \hat{f}_{i+1} + t \mathrm{e}^{\imath \pi N} \hat{f}_{L}^{\dagger} \hat{f}_{1} \hat{P} + \mathrm{H.c}.
\end{equation}
For the antiperiodic boundary condition an analogous calcuation can be done to find the Hamiltonian:
\begin{equation}
  \label{eq:app:Ham-ferm-apbc}
  H = -t \sum\limits_{i=1}^{L-1} \hat{f}_{i}^{\dagger} \hat{f}_{i+1} - t \mathrm{e}^{\imath \pi N} \hat{f}_{L}^{\dagger} \hat{f}_{1} \hat{P} + \mathrm{H.c}.
\end{equation}

\section{Calculation of ground state energy}
\label{app:groundenergy}
\renewcommand{\thesubsubsection}{}
We calculate the ground state energies for all eight possible cases shown in Tab. \ref{tab:Ground-ener-list}. The ground state energies are found by summing all the \emph{N} lowest lying energy levels. In Eqs. \eqref{eq:eigen-energy-even}, \eqref{eq:eigen-energy-odd}, \eqref{eq:eigen-energy-apbc-even}, \eqref{eq:eigen-energy-apbc-odd}, \eqref{eq:eigen-energy-b-field-even}, \eqref{eq:eigen-energy-b-field-odd}, \eqref{eq:eigen-energy-b-field-apbc-even} and \eqref{eq:eigen-energy-b-field-apbc-odd} the $k$-th mode energies have a general form involving \emph{cosine}. Therefore, we first find a general summation formula over \emph{cosine} terms, and later use that formula to calculate the ground state energies.
\begin{table}
  \begin{ruledtabular}
    \begin{tabular}{c c c c c c}
      Case &N   &Boundary cond.    &Mag. field &E$_{g}$ &E$_{k}$\\
      \hline
      I &Odd &Periodic        &No         &\eqref{eq:app:Eg:odd-N-PBC-wo-mag} &\eqref{eq:eigen-energy-odd} \\
      II &Odd &Antiperiodic    &No         &\eqref{eq:app:Eg:odd-N-APBC-wo-mag} &\eqref{eq:eigen-energy-apbc-odd} \\
      III &Odd &Periodic        &Yes         &\eqref{eq:app:Eg:odd-N-PBC-with-mag} &\eqref{eq:eigen-energy-b-field-odd} \\
      IV &Odd &Antiperiodic    &Yes         &\eqref{eq:app:Eg:odd-N-APBC-with-mag} &\eqref{eq:eigen-energy-b-field-apbc-odd} \\
      V &Even &Periodic        &No         &\eqref{eq:app:even-n-pbc-nomag} &\eqref{eq:eigen-energy-even} \\
      VI &Even &Antiperiodic    &No         &\eqref{eq:app:even-n-apbc-nomag} &\eqref{eq:eigen-energy-apbc-even} \\
      VII&Even &Periodic        &Yes         &\eqref{eq:app:even-n-pbc-mag} &\eqref{eq:eigen-energy-b-field-even} \\
      VIII &Even &Antiperiodic    &Yes         &\eqref{eq:app:even-n-apbc-mag} &\eqref{eq:eigen-energy-b-field-apbc-even}\\
    \end{tabular}
  \end{ruledtabular}
  \caption{\label{tab:Ground-ener-list} List of the eight possible cases and their corresponding equation No. for the ground state energies. Here, the column <<N>> represents whether even or odd number of particles are present in the ring. The column <<E$_{g}$>> gives the equation No. of the ground state energy. The actual expressions and derivations of these ground state energies are given in App. \ref{app:groundenergy}. The column <<E$_{k}$>> gives the equation No. of the $k$-th mode energy.}
\end{table}

The single mode energies $E_{k}$ are in the general form:
\begin{equation}
  \label{eq:app:cos}
  E_{k} = -2t \cos \left( a + k \cdot d \right).
\end{equation}
 Here $a$ is the constant phase, and $k \cdot d$ is the phase in arithmetic progression. We will first find a general summation formula of Eq. (\ref{eq:app:cos}). The summation over \emph{cosine} terms can be found easily by representing them as the real part of exponential, $\Re \{\exp [i (a + k \cdot d)] \}$, which transforms the summation to the geometric series:
  \begin{equation}
    \label{eq:app:sum-geom-series}
    \begin{aligned}
      \sum\limits_{k=0}^{n-1} E_{k}&=
      \sum\limits_{k=0}^{n-1} \cos (a+ k \cdot d) \\ &= \sum\limits_{k=0}^{n-1} \Re \left\{\mathrm{e}^{\imath (a + k \cdot d)}\right\}\\ &=\Re \left\{\mathrm{e}^{\imath a} \sum\limits_{k=0}^{n-1} \left( \mathrm{e}^{\imath \cdot d}\right)^{k} \right\}\\ &= \Re \left\{ \mathrm{e}^{\imath a} \left( \frac{1-\mathrm{e}^{\imath d n}}{1 - \mathrm{e}^{\imath d}} \right)\right\}\\ &= \Re \left\{ \mathrm{e}^{\imath a} \left[ \frac{\mathrm{e}^{\imath d n/2}}{\mathrm{e}^{\imath d/2}} \left( \frac{e^{-\imath d n/2} - e^{\imath d n/2}}{e^{-\imath d/2} - e^{\imath d/2}} \right) \right] \right\}\\ &= \Re \left\{ \mathrm{e}^{\imath [a + (n-1)d/2]} \left[ \frac{\sin \left( nd/2 \right)}{\sin \left( d/2 \right)} \right] \right\}\\ &= \left[ \frac{\sin \left( nd/2 \right)}{\sin \left( d/2 \right)} \right] \times \cos \left[ \frac{2a + (n-1)d}{2} \right].
    \end{aligned}
  \end{equation}
Analogously the summation with negative $k$ is:
\begin{equation}
  \label{eq:app:negative-m}
  \begin{aligned}
    \sum\limits_{k=0}^{n-1}& \cos (a- k \cdot d) \\ &= \left[ \frac{\sin \left( nd/2 \right)}{\sin \left( d/2 \right)} \right] \times \cos \left[ \frac{2a + (1-n)d}{2} \right].
    \end{aligned}
\end{equation}

\subsection{Ground sate energy for odd $N$}
\label{sec:app:ground-sate-energy-odd-n}
We first calculate the ground state energy for odd $N$. For odd $N$, the quantum number $k$ takes the values $-(N-1)/2, -[(N-1)/2] + 1, \dots, [(N-1)/2] - 1, (N-1)/2$. Equivalent representation of this summation is:
\begin{equation}
  \label{eq:app:sum-odd-N}
  \begin{aligned}
    \sum\limits_{-(N-1)/2}^{(N-1)/2} \mathrm{E}_k &= \sum\limits_{0}^{(N-1)/2} \mathrm{E}_k + \sum\limits_{-1}^{-(N-1)/2} \mathrm{E}_k \\ &= \sum\limits_{0}^{(N-1)/2} \mathrm{E}_k + \sum\limits_{0}^{-(N-1)/2} \mathrm{E}_k - \mathrm{E}_0\\ &= \sum\limits_{0}^{(N-1)/2} \mathrm{E}_k + \sum\limits_{0}^{(N-1)/2} \mathrm{E}_{-k} - \mathrm{E}_{0}.
  \end{aligned}  
\end{equation}
The ground state energy for four cases corresponding to odd \emph{N} are calculated below.

Case-I: Odd $N$, Periodic Boundary Condition, Without Magnetic Field.
In Eq. \eqref{eq:eigen-energy-odd} the $k$-th mode energy for odd \emph{N} with periodic boundary condition and without magnetic field is given. Comparing Eq. \eqref{eq:eigen-energy-odd} and Eq. \eqref{eq:app:cos} we find:
\begin{equation}
  \label{eq:app:comp-1}
  \begin{aligned}
    a \mapsto \frac{2 \pi}{L} \left( \frac{p_{\nu}}{N_{\nu}}\right); \: d \mapsto \frac{2\pi}{L}.
  \end{aligned}
\end{equation}
Substituting these values into Eq. \eqref{eq:app:sum-geom-series}, and summing over $n= (N+1)/2$ states we find:
\begin{equation}
  \label{eq:app:oddN-period-wo-mag-1}
  \begin{aligned}
    \sum\limits_{k=0}^{(N-1)/2}E_{k} =&  -2t\frac{ \sin \left[(N+1)\pi/2L \right]}{\sin (\pi/L)}\\ &\quad \times \cos \left[ \frac{\frac{4\pi}{L}\left( \frac{p_{\nu}}{N_{\nu}}  \right) + \frac{(N-1)\pi}{L}}{2} \right].
    \end{aligned}
\end{equation}
Similarly, substituting Eq. \eqref{eq:app:comp-1} into Eq. \eqref{eq:app:negative-m} , and summing over $n= (N+1)/2$ states we find:
\begin{equation}
  \label{eq:app:oddN-period-wo-mag-2}
  \begin{aligned}
    \sum\limits_{k=0}^{(N-1)/2}E_{-k} = &-2t\frac{ \sin \left[ (N+1)\pi/2L \right]}{\sin (\pi/L)}\\ &\quad \times \cos \left[ \frac{\frac{4\pi}{L}\left( \frac{p_{\nu}}{N_{\nu}}  \right) + \frac{(1-N)\pi}{L}}{2} \right].
    \end{aligned}
\end{equation}
The $k=0$-th mode energy will be:
\begin{equation}
  \label{eq:app:oddN-period-wo-mag-3}
  E_{0} = -2t \cos \left[ \frac{2 \pi}{L} \left( \frac{p_{\nu}}{N_{\nu}} \right) \right].
\end{equation}
Substituting Eqs. \eqref{eq:app:oddN-period-wo-mag-1}, \eqref{eq:app:oddN-period-wo-mag-2} and (\ref{eq:app:oddN-period-wo-mag-3}) in \eqref{eq:app:sum-odd-N} we find the ground state energy:
  \begin{equation}
    \label{eq:app:Eg:odd-N-PBC-wo-mag}
    \begin{aligned}
      &E_{PBC, \: g} =\\ &-2t\frac{\sin \left[ (N+1)\pi/2L \right]}{\sin (\pi/L)} \cos \left[ \frac{\frac{4\pi}{L}\left( \frac{p_{\nu}}{N_{\nu}} \right) + \frac{(N-1)\pi}{L}}{2} \right]\\ & -2t\frac{ \sin\left[ (N+1)\pi/2L \right]}{\sin (\pi/L)} \cos \left[ \frac{\frac{4\pi}{L}\left( \frac{p_{\nu}}{N_{\nu}}  \right) + \frac{(1-N)\pi}{L}}{2} \right]\\ &+ 2t \cos \left[ \frac{2 \pi}{L} \left( \frac{p_{\nu}}{N_{\nu}}  \right) \right].
      \end{aligned}
    \end{equation}

Case-II: Odd $N$, Antiperiodic Boundary Condition, Without Magnetic Field.
In Eq. (\ref{eq:eigen-energy-apbc-odd}) the $k$-th mode energy for odd \emph{N} with antiperiodic boundary condition and without magnetic field is given. Comparing Eq. \eqref{eq:eigen-energy-apbc-odd} and Eq. \eqref{eq:app:cos} we find:
\begin{equation}
  \label{eq:app:comp-2}
  \begin{aligned}
    a \mapsto \frac{2 \pi}{L} \left( \frac{p_{\nu}}{N_{\nu}} + \frac{1}{2}\right); \: d \mapsto \frac{2\pi}{L}.
  \end{aligned}
\end{equation}
Repeating the procedure analogous to Eq. \eqref{eq:app:oddN-period-wo-mag-1}, \eqref{eq:app:oddN-period-wo-mag-2} and \eqref{eq:app:oddN-period-wo-mag-3} we find:
\begin{equation}
  \label{eq:app:Eg:odd-N-APBC-wo-mag}
  \begin{aligned}
    &E_{APBC, \: g} =\\ &-2t\frac{\sin \left[ (N+1)\pi/2L \right]}{\sin (\pi/L)} \cos \left[ \frac{\frac{4\pi}{L}\left( \frac{p_{\nu}}{N_{\nu}} + \frac{1}{2} \right) + \frac{(N-1)\pi}{L}}{2} \right]\\ &-2t\frac{\sin \left[ (N+1)\pi/2L \right]}{\sin (\pi/L)} \cos \left[ \frac{\frac{4\pi}{L}\left( \frac{p_{\nu}}{N_{\nu}} + \frac{1}{2} \right) + \frac{(1-N)\pi}{L}}{2} \right] \\  &+ 2t \cos \left[ \frac{2 \pi}{L} \left( \frac{p_{\nu}}{N_{\nu}} + \frac{1}{2}\right) \right].
    \end{aligned}
\end{equation}

Case-III: Odd $N$, Periodic Boundary Condition, With Magnetic Field.
$k$-th mode energy for periodic boundary condition with magnetic field contains two extra terms: (i) energy contribution due to \emph{Zeeman effect} ($Z_{\nu}$), (ii) phase contribution due to Aharanov-Bohm effect ($2 \pi \Phi_{B}/(\Phi_{0}L)$) as shown in Eq. (\ref{eq:eigen-energy-b-field-odd}). Hence in Eq. \eqref{eq:app:Eg:odd-N-PBC-wo-mag} if we just add an extra phase and \emph{Zeeman energy} we can directly get the ground state energy:
\begin{equation}
  \label{eq:app:Eg:odd-N-PBC-with-mag}
  \begin{aligned}
    &E^{(B)}_{PBC, \: g} =\\ &-2t\frac{\sin\left[  (N+1)\pi/2L \right]}{\sin (\pi/L)} \cos \left[ \frac{\frac{4\pi}{L}\left( \frac{p_{\nu}}{N_{\nu}} + \frac{\Phi_{B}}{\Phi_0} \right) + \frac{(N-1)\pi}{L}}{2} \right]\\  &-2t\frac{ \sin\left[ (N+1)\pi/2L \right]}{\sin (\pi/L)} \cos \left[ \frac{\frac{4\pi}{L}\left( \frac{p_{\nu}}{N_{\nu}}  + \frac{\Phi_{B}}{\Phi_0} \right) + \frac{(1-N)\pi}{L}}{2} \right]\\ & +2t \cos \left[ \frac{2 \pi}{L} \left( \frac{p_{\nu}}{N_{\nu}}  + \frac{\Phi_{B}}{\Phi_0}\right) \right] + Z_{\nu}.
    \end{aligned}
\end{equation}

Case-IV: Odd $N$, Antiperiodic Boundary Condition, With Magnetic Field.
\label{sec:app:odd-n-antiperiodic-1-with-mag}
As in Eq. \eqref{eq:app:Eg:odd-N-PBC-with-mag}, we can find the ground state energy directly from Eq. \eqref{eq:app:Eg:odd-N-APBC-wo-mag} by adding a phase $2 \pi \Phi_{B}/(\Phi_{0}L)$ and \emph{Zeeman energy} $Z_{\nu}$. The ground state energy is:
\begin{equation}
  \label{eq:app:Eg:odd-N-APBC-with-mag}
  \begin{aligned}
    &E^{(B)}_{PBC, \: g} =\\ &-2t\frac{ \sin\left[ (N+1)\pi/2L \right]}{\sin (\pi/L)}  \cos \left[ \frac{\frac{4\pi}{L}\left( \frac{p_{\nu}}{N_{\nu}} + \frac{1}{2} + \frac{\Phi_{B}}{\Phi_0} \right) + \frac{(N-1)\pi}{L}}{2} \right]\\ &-2t\frac{\sin\left[  (N+1)\pi/2L \right]}{\sin (\pi/L)} \cos \left[ \frac{\frac{4\pi}{L}\left( \frac{p_{\nu}}{N_{\nu}} + \frac{1}{2} + \frac{\Phi_{B}}{\Phi_0} \right) + \frac{(1-N)\pi}{L}}{2} \right] \\ & +2t \cos \left[ \frac{2 \pi}{L} \left( \frac{p_{\nu}}{N_{\nu}}  + \frac{1}{2} + \frac{\Phi_{B}}{\Phi_0}\right) \right] + Z_{\nu}.
    \end{aligned}  
\end{equation}

\subsection{Ground sate energy for even $N$}
\label{sec:app:ground-sate-energy-odd-n}
Ground state energy for even $N$ is found by summing the energies over $-(N/2), -(N/2)+1, \dots, -1,1,2,\dots, (N/2)$ energy levels. Equivalent representation of this summation is:
\begin{equation}
  \label{eq:app:sum-even-N}
  \begin{aligned}
    \sum\limits_{-N/2}^{(N/2)} \mathrm{E}_k &= \sum\limits_{0}^{(N/2)} \mathrm{E}_k + \sum\limits_{-1}^{-N/2} \mathrm{E}_k\\ &= \sum\limits_{0}^{(N/2)} \mathrm{E}_k + \sum\limits_{0}^{-N/2} \mathrm{E}_k - 2\mathrm{E}_0\\ &= \sum\limits_{0}^{(N/2)-1} \mathrm{E}_k + \sum\limits_{0}^{N/2} \mathrm{E}_{-k} - 2\mathrm{E}_{0}.
  \end{aligned}  
\end{equation}
As in odd $N$ case, for even $N$ also arises four different cases. All these four cases are calculated below.

Case-V: Even $N$, Periodic Boundary Condition, Without Magnetic Field.
In Eq. (\ref{eq:eigen-energy-even}) the $k$-th mode energy for even \emph{N} with antiperiodic boundary condition and without magnetic field is given. Comparing Eq. \eqref{eq:eigen-energy-even} for even $N$ and Eq. \eqref{eq:app:cos} we find:
\begin{equation}
  \label{eq:app:comp-2}
  \begin{aligned}
    a \mapsto \frac{2 \pi}{L} \left( \frac{p_{\nu}}{N_{\nu}} + \frac{1}{2} \right); \: d \mapsto \frac{2\pi}{L}.
  \end{aligned}
\end{equation}
It should be noted that now we sum the positive $k$ values from $1$ to $(N/2)$, and negative $k$ values from $-1$ to $-(N/2)$. Repeating the procedure analogous to Eq. \eqref{eq:app:oddN-period-wo-mag-1}, \eqref{eq:app:oddN-period-wo-mag-2} and \eqref{eq:app:oddN-period-wo-mag-3} we find:
\begin{equation}
  \label{eq:app:even-n-pbc-nomag}
  \begin{aligned}
    &E_{PBC,g} =\\ &-2t \frac{\sin \left[ ((N+2) \pi)/2L \right]}{\sin (\pi/L)} \: \cos \left[ \frac{\frac{4 \pi}{L} \left(\frac{p_{\nu}}{N_{\nu}} + \frac{1}{2} \right) + \frac{(N) \pi}{L}}{2}\right]\\ & -2t \frac{\sin \left[ (N+2)\pi/2L \right]}{\sin (\pi/L)}\: \cos \left[ \frac{\frac{4 \pi}{L} \left(\frac{p_{\nu}}{N_{\nu}} + \frac{1}{2} \right) - \frac{N \pi}{L}}{2} \right]\\
    &+4t \cos \left[ \frac{2\pi}{L} \left( \frac{p_{\nu}}{N_{\nu}} + \frac{1}{2} \right) \right].
  \end{aligned}
\end{equation}

Case-VI: Even $N$, Antiperiodic Boundary Condition, Without Magnetic Field.
In Eq. (\ref{eq:eigen-energy-apbc-even}) the $k$-th mode energy for even \emph{N} with antiperiodic boundary condition and without magnetic field is given. Comparing Eq. \eqref{eq:eigen-energy-apbc-even} for even $N$ and Eq. \eqref{eq:app:cos} we find:
\begin{equation}
  \label{eq:app:comp-even-n-3}
  \begin{aligned}
    a \mapsto \frac{2 \pi}{L} \left( \frac{p_{\nu}}{N_{\nu}} \right); \: d \mapsto \frac{2\pi}{L}.
  \end{aligned}
\end{equation}
The ground state energy can be found easily from Eq. \eqref{eq:app:even-n-pbc-nomag} by replacing the phase $a$ by new values. The ground state energy is:
\begin{equation}
  \label{eq:app:even-n-apbc-nomag}
  \begin{aligned}
    &E_{APBC,g} =\\ &-2t \frac{\sin \left[ ((N+2) \pi)/2L \right]}{\sin (\pi/L)} \: \cos \left[ \frac{\frac{4 \pi}{L} \left(\frac{p_{\nu}}{N_{\nu}}\right) + \frac{(N) \pi}{L}}{2}\right]\\ & -2t \frac{\sin \left[ (N+2)\pi/2L \right]}{\sin (\pi/L)}\: \cos \left[ \frac{\frac{4 \pi}{L} \left(\frac{p_{\nu}}{N_{\nu}} \right) - \frac{N \pi}{L}}{2} \right]\\
    &+4t \cos \left[ \frac{2\pi}{L} \left( \frac{p_{\nu}}{N_{\nu}} \right) \right].
  \end{aligned}
\end{equation}

Case-VII: Even $N$, Periodic Boundary Condition, With Magnetic Field.
Ground sate for this case can easily found by adding the \emph{Aharanov-Bohm phase}, and the \emph{Zeeman energy} in Eq. \eqref{eq:app:even-n-pbc-nomag}. The ground state energy is:
\begin{equation}
  \label{eq:app:even-n-pbc-mag}
  \begin{aligned}
    &E^{(B)}_{PBC,g} =\\ &-2t \frac{\sin \left[ ((N+2) \pi)/2L \right]}{\sin (\pi/L)} \: \cos \left[ \frac{\frac{4 \pi}{L} \left(\frac{p_{\nu}}{N_{\nu}} + \frac{1}{2} + \frac{\Phi_{B}}{\Phi_0}\right) + \frac{(N) \pi}{L}}{2}\right]\\ & -2t \frac{\sin \left[ (N+2)\pi/2L \right]}{\sin (\pi/L)}\: \cos \left[ \frac{\frac{4 \pi}{L} \left(\frac{p_{\nu}}{N_{\nu}}+ \frac{1}{2} + \frac{\Phi_{B}}{\Phi_0}\right) - \frac{N \pi}{L}}{2} \right]\\
    &+4t \cos \left[ \frac{2\pi}{L} \left( \frac{p_{\nu}}{N_{\nu}} + \frac{1}{2} + \frac{\Phi_{B}}{\Phi_0} \right) \right] + Z_{\nu}.
  \end{aligned}
\end{equation}

Case-VIII: Even $N$, Antiperiodic Boundary Condition, With Magnetic Field.
Ground sate for this case can easily found by adding the \emph{Aharanov-Bohm phase}, and the \emph{Zeeman energy} in Eq. \eqref{eq:app:even-n-apbc-nomag}. The ground state energy is:
\begin{equation}
  \label{eq:app:even-n-apbc-mag}
  \begin{aligned}
    &E^{(B)}_{APBC,g} =\\ &-2t \frac{\sin \left[ ((N+2) \pi)/2L \right]}{\sin (\pi/L)} \: \cos \left[ \frac{\frac{4 \pi}{L} \left(\frac{p_{\nu}}{N_{\nu}} + \frac{\Phi_{B}}{\Phi_0}\right) + \frac{(N) \pi}{L} \frac{\pi}{L}}{2}\right]\\ & -2t \frac{\sin \left[ (N+2)\pi/2L \right]}{\sin (\pi/L)}\: \cos \left[ \frac{\frac{4 \pi}{L} \left(\frac{p_{\nu}}{N_{\nu}}  + \frac{\Phi_{B}}{\Phi_0}\right) - \frac{N \pi}{L}}{2} \right]\\
    &+4t \cos \left[ \frac{2\pi}{L} \left( \frac{p_{\nu}}{N_{\nu}}  + \frac{\Phi_{B}}{\Phi_0} \right) \right].
  \end{aligned}
\end{equation}